\documentclass{elsart}
\usepackage{graphicx,amsmath,amssymb,bm,feynmp}
\begin{document}
\begin{fmffile}{feyn}

\begin{frontmatter}
\title{Low Momentum Nucleon-Nucleon Interaction and Fermi Liquid Theory}
\author{Achim Schwenk$^{(a)}$\thanksref{AS}},
\author{Gerald E. Brown$^{(a)}$\thanksref{GEB}}, and
\author{Bengt Friman$^{(b)}$\thanksref{BF}}
\address{$^{(a)}$Department of Physics and Astronomy,
State University of New York,\\
Stony Brook, N.Y. 11794-3800, U.S.A.\\
$^{(b)}$Gesellschaft f\"ur Schwerionenforschung, Planckstr. 1, 64291
Darmstadt, Germany}

\thanks[AS]{E-mail: aschwenk@nuclear.physics.sunysb.edu}
\thanks[GEB]{E-mail: popenoe@nuclear.physics.sunysb.edu}
\thanks[BF]{E-mail: b.friman@gsi.de}

\begin{abstract}
We use the induced interaction of Babu and Brown to derive two
novel relations between the quasiparticle interaction in nuclear matter
and the unique low momentum nucleon-nucleon interaction $V_{\text{low
k}}$ in vacuum. These relations provide two independent constraints on
the Fermi liquid parameters of nuclear matter.
We derive the full renormalization group equations in the
particle-hole channels from the induced interaction. The new
constraints, together with the Pauli principle sum rules, define
four combinations of Fermi liquid parameters that are invariant under
the renormalization group flow. Using empirical values for the
spin-independent Fermi liquid parameters, we are able to compute the
major spin-dependent ones by imposing the new constraints and the
Pauli principle sum rules. The effects of tensor forces are discussed.

\vspace{0.5cm}

\end{abstract}
\vspace*{1.5cm}
{\small We dedicate this paper to the memory of\\ Sven-Olof B\"ackman.}
\vspace*{.5cm}
\end{frontmatter}

\newpage

\section{Introduction}

This work was motivated by the results of Bogner, Kuo and
Corragio~\cite{Vlowk}, who have constructed a low momentum
nucleon-nucleon potential $V_{\text{low k}}$ using folded-diagram
techniques. The starting point of their procedure is a realistic
nucleon-nucleon interaction, which is reduced to a low momentum potential by
integrating out relative momenta higher than a cutoff $\Lambda$, in
the sense of the renormalization group (RG)~\cite{Vlowkflow}. The hard
momenta larger than $\Lambda$ renormalize $V_{\text{low k}}$, such that
the low momentum half-on-shell $T$ matrix and bound state properties
of the underlying theory remain unchanged. Consequently, the physics
at relative momenta smaller than $\Lambda$ is preserved.

Bogner {\it et al.} find that various, very different bare
interactions, such as the Paris, Bonn, and Argonne
potential and a chiral effective
field theory model, flow to the same $V_{\text{low k}}$ for $\Lambda
\lesssim 2 \; \text{fm}^{-1}$~\cite{Vlowkflow}. All
the nuclear force models are constructed to fit the experimentally
available nucleon-nucleon phase shifts up to momenta $k \sim 2 \;
\text{fm}^{-1}$. However, they differ substantially in their treatment
of the short range parts of the interaction, since these effects cannot
be pinned down uniquely by the scattering data. Therefore the work of
Bogner {\it et al.} demonstrates that one can isolate the physics of the
nucleons at low momenta from the effects probed by high momenta and
in this way obtain a unique low momentum nucleon-nucleon
potential $V_{\text{low k}}$. When one compares the low momentum
part of the bare potentials with $V_{\text{low k}}$, one observes
that for reasonable values of the cutoff
the main effect of the RG decimation to a unique $V_{\text{low
k}}$ is a constant shift in momentum space corresponding to a delta
function in coordinate space~\footnote{Due to the cutoff employed,
the constant shift within the model space corresponds to a smeared
delta function.}. This is in keeping with the ideas of effective
field theory, where one projects $V_{\text{low k}}$ on one and two
pion exchange terms plus contact terms, the latter resulting from
the exchange of the heavy mesons. The non-pionic contact term 
contributions flow to ``fixed point'' values for $\Lambda \lesssim 2 \;
\text{fm}^{-1}$. Therefore, the most important feature of the
unique low momentum interaction is its value at zero initial and
final relative momenta $V_{\text{low k}} (0,0)$, since it directly
incorporates the largest effect of the RG decimation -- the removal
of the model dependent short range core by a smeared delta function.

Moreover, it is seen~\cite{Vlowkflow} that in the $^1 S_0$ channel,
$V_{\text{low k}} (0,0)$
is almost independent of the cutoff $\Lambda$ for
$1\;\text{fm}^{-1}\lesssim\Lambda\lesssim 3 \;\text{fm}^{-1}$,
while in the $^3 S_1$ channel only a weak linear dependence on
$\Lambda$ remains in the same range of momenta. For $\Lambda$ in this
momentum range, the contribution of the short range repulsion,
which is peaked around approximately $4\;\text{fm}^{-1}$~\cite{Bethe},
is already integrated out, while the common one pion exchange long
range tail remains
basically unchanged until $\Lambda \sim m_\pi$.
The residual dependence on the cutoff in the $^3 S_1$ channel is
due to higher order tensor contributions, which are peaked at an
intermediate momentum transfer of approximately $2
\; \text{fm}^{-1}$~\cite{UTNMF}. The weak cutoff dependence of
$V_{\text{low k}}$ around $\Lambda \sim 2 \;
\text{fm}^{-1}$ is characteristic of effective field theories, where the
dependence on the cutoff is expected to be weak, provided the relevant
degrees of freedom -- here nucleons and pions -- are kept explicitly.
The separation of scales implied by the exchanged meson masses is
complicated, however, by the higher order tensor interactions.

Diagrammatically $V_{\text{low k}}$ sums all ladders with bare
potential vertices and intermediate momenta greater than the
cutoff. Subsequently, the energy dependence of the ladder sum is
removed in order to obtain an energy independent $V_{\text{low
k}}$. This is achieved by means of folding, which can be regarded
as averaging over the energy dependent effective interaction
weighted by the low momentum components of the low energy
scattering states. Therefore, it is intuitive to use $V_{\text{low
k}}$ for $\Lambda = k_{\text{F}}$ as the Brueckner $G$ matrix. This
identification is approximative, since self energy insertions and
the dependence on the center of mass momentum are ignored in
$V_{\text{low k}}$. However, it has been argued that the self energy
insertions, which must be evaluated off-shell, are small~\cite{UTNMF}.
Hence, we expect that $V_{\text{low k}}$ reproduces the $G$ matrix
reasonably well. Furthermore, Bogner {\it et al.}~\cite{18O} argue that
$V_{\text{low k}}$ may be used directly as a shell model effective
interaction instead of the Brueckner $G$ matrix, since
$V_{\text{low k}}$ includes the effects of the repulsive core
and is generally smooth. They find very good agreement for the low
lying states of core nuclei with two valence nucleons such as
$^{18}$O and $^{134}$Te.

Our second motivation is the work of Birse {\it et al.} on the
Wilsonian renormalization group treatment of two-body
scattering~\cite{RG2body,RG2bodylong}, where the existence of a
unique low momentum potential is addressed. By demanding that the
physical $T$ matrix be independent of the cutoff~\footnote{Their
analysis was carried out for the reaction matrix, but a similar
analysis holds for the T matrix. In~\cite{VlowkRG} the RG equation
for $V_{\text{low k}}$ is derived from the Lippmann-Schwinger
equation for the half-on-shell T-matrix and it is shown that the
same flow equation can be equivalently obtained from the
Kuo-Lee-Ratcliff folded diagram series and the Lee-Suzuki
similarity transformation.}, they obtain a RG flow equation for the
effective potential. After rescaling all dimensionful quantities
with the cutoff, they find a trivial fixed point corresponding to
zero scattering length and a nontrivial one corresponding to an
infinite scattering length. The expansion around the nontrivial
fixed point yields the effective range expansion. This demonstrates
that the s-wave nucleon-nucleon potential, where the scattering
length is large, must lie in the vicinity of the nontrivial fixed
point. It would be of interest to clarify the role of this fixed point
structure in the RG flow to $V_{\text{low k}}$ and in particular
whether this can be used to understand why a {\it unique} potential 
is obtained already for $\Lambda \lesssim 2 \; \text{fm}^{-1}$.

In normal Fermi systems, the low momentum quasiparticle interaction, which is
characterized by the Fermi liquid parameters, is determined by
a RG fixed point. In this paper we derive a
relation between the Fermi liquid parameters of nuclear matter and
the s-wave low momentum nucleon-nucleon interaction $V_{\text{low k}} (0,0)$
at $\Lambda = k_{\text{F}}$. This relation connects the fixed point of
the quasiparticle interaction to $V_{\text{low k}}$ in the region
where it depends only weakly on the cutoff. The existence of such a relation
is supported by the success of the model space calculations of Bogner
{\it et al.}~\cite{18O}, where $V_{\text{low k}}$ is used as the shell
model effective interaction. These calculations are in spirit very
similar to Fermi liquid theory. In both cases one uses an empirical
single-particle spectrum and the energy is measured with respect to a
filled Fermi sea. In the case of $^{18}$O, the zero of the energy
corresponds to the ground state of $^{16}$O.

We start by giving a brief introduction of Landau's theory of
normal Fermi liquids. We then review the induced interaction
introduced by Babu and Brown~\cite{BB}, which will be used to
derive the two new constraints. We give a diagrammatically
motivated heuristic derivation of the induced interaction, which
demonstrates that the induced interaction generates
the complete particle-hole parquet for the scattering amplitude,
i.e. all fermionic planar diagrams except for particle-particle
loops. The latter should be included in the driving term. We then
derive two new constraints that relate the Fermi liquid parameters
to the low momentum nucleon-nucleon interaction $V_{\text{low k}}$,
by solving the integral equation for the scattering amplitude and
the induced interaction in a particular limit simultaneously.
Making contact with the RG approach to Fermi
liquid theory, we derive the coupled RG equations for the
particle-hole channels from the induced interaction. Within the
particle-hole parquet, the particular combinations 
of Fermi liquid parameters that appears in these constraints, as
well as the Pauli principle sum rules, are invariant under the (in
medium) RG flow towards the Fermi surface. Using empirical values
for the spin-independent Fermi liquid parameters, we are able to
compute the major spin-dependent parameters by imposing the new
constraints and the Pauli principle sum rules. Finally, we include
tensor interactions in the constraints and demonstrate the
necessity of a self-consistent treatment within the induced
interaction.

\section{Fermi Liquid Theory}

Fermi liquid theory was invented by Landau~\cite{Landau} to
describe strongly interacting normal Fermi systems at low
temperatures. Landau introduced the quasiparticle concept to
describe the elementary excitations of the interacting system. For
low excitation energies, the corresponding quasiparticles are long
lived and in a sense weakly interacting. One can think of the
ground state of the system as a filled Fermi sea of quasiparticles,
while quasiparticles above and quasiholes below the Fermi surface
correspond to low-lying excited states. The quasiparticles can be
thought of as free particles dressed by the
interactions with the many-body medium.

When quasiparticles or quasiholes are added to the interacting
ground state, the energy of the system is changed by
\begin{equation}
\delta E = \sum_{{\mathbf p} \sigma} \epsilon^{(0)}_{{\mathbf p}}
\delta n_{{\mathbf p} \sigma} + \frac{1}{2 V} \sum_{{\mathbf p} \sigma
, {\mathbf p^\prime} \sigma^\prime} f_{\sigma, \sigma^\prime}
({\mathbf p}, {\mathbf p^\prime}) \delta n_{{\mathbf p} \sigma}
\delta n_{{\mathbf p^\prime} \sigma^\prime} + {\mathcal O}(\delta n^3),
\label{deltae}
\end{equation}
where V is the volume of the system, $\delta n_{{\mathbf p}
\sigma}$ the change in the quasiparticle occupation number and
$\epsilon^{(0)}_{{\mathbf p}} - \mu= v_{\text{F}} (p -
k_{\text{F}})$ the quasiparticle energy expanded
around the Fermi surface. The Fermi momentum is denoted by
$k_{\text{F}}$, the Fermi velocity by $v_{\text{F}}$ and the
chemical potential by $\mu$. The quasiparticle lifetime in normal
Fermi systems at zero temperature, is very large close to the Fermi
surface ($\tau\sim (p-k_{\text{F}})^{-2}$). Consequently, the
quasiparticle concept is useful for describing long wavelength
excitations, where the corresponding quasiparticles are restricted
to momenta $|{\mathbf p}| \approx k_{\text{F}}$.  When studying
such excitations, one can set $|{\mathbf p}|=|{\mathbf
p^\prime}|=k_{\text{F}}$ in the effective interaction
$f_{\sigma,\sigma^\prime}({\mathbf p},{\mathbf p^\prime})$. In a
rotationally invariant system, the only remaining spatial variable
of $f$ is then the angle $\theta$ between ${\mathbf p}$ and
${\mathbf p^\prime}$. The dependence of
$f_{\sigma,\sigma^\prime}({\mathbf p},{\mathbf p^\prime})$ on this
angle reflects the non-locality of the quasiparticle interaction.

It follows from Eq. (\ref{deltae}) that the effective interaction is
obtained from the energy by varying twice with respect to the
quasiparticle occupation number. An illustrative example is the
Hartree-Fock approximation, where one finds that the Landau $f$ function
is simply given by the direct and exchange terms of the bare interaction:
\begin{equation}
\frac{\delta E^{\text{HF}}}{\delta n_{{\mathbf p} \sigma} \delta
n_{{\mathbf p^\prime} \sigma^\prime}} = \frac{f_{\sigma,
\sigma^\prime}^{\text{HF}} ({\mathbf p}, {\mathbf p^\prime})}{V} = \langle
{\mathbf p} \sigma, {\mathbf p^\prime} \sigma^\prime | V | {\mathbf p}
\sigma, {\mathbf p^\prime} \sigma^\prime \rangle - \langle {\mathbf p}
\sigma, {\mathbf p^\prime} \sigma^\prime | V | {\mathbf p^\prime}
\sigma^\prime, {\mathbf p} \sigma \rangle.
\label{HF}
\end{equation}
As in effective field theories, the functional form of the
spin- and isospin-dependence of the
Landau function is determined by
the symmetries of the system only -- in the case of symmetric nuclear
matter these are invariance under spin and isospin
rotations~\footnote{When tensor forces are considered the
quasiparticle interaction is not invariant under rotations in spin space, but
under combined spin and spatial rotations.}. The dependence of $f$
on the angle $\theta$ is expanded in Legendre polynomials:
\begin{multline}
\hspace{-0.3cm} f(\theta) = \frac{1}{N(0)} {\mathcal F}(\theta) =
\frac{1}{N(0)} \sum_{l}
\biggl( F_l + F_l^{\prime} \:
{\bm \tau} \cdot {\bm \tau^{\prime}} + G_l \: {\bm \sigma} \cdot {\bm
\sigma^{\prime}} + G^{\prime}_l \: {\bm \tau} \cdot {\bm
\tau^{\prime}} \: {\bm \sigma} \cdot {\bm \sigma^{\prime}} \biggr) P_l
(\cos \theta) \\[0.5mm]
+  \frac{1}{N(0)} \frac{({\mathbf p} - {\mathbf
p^\prime})^2}{k_{\text{F}}^2} \: S_{12}({\mathbf p} -
{\mathbf p^\prime}) \sum_l \biggl( H_l +
H^\prime_l \: {\bm \tau} \cdot {\bm \tau^{\prime}} \biggr) P_l (\cos
\theta) + {\mathcal O}(A^{-1/3}).
\label{LF}
\end{multline}
Here ${\bm \sigma}$ and ${\bm \tau}$ are spin and isospin operators
respectively, $S_{12}({\mathbf k}) = 3 \: {\bm \sigma} \cdot
\widehat{{\mathbf k}} \; {\bm \sigma^{\prime}} \cdot \widehat{{\mathbf k}} -
{\bm \sigma} \cdot {\bm \sigma^{\prime}}$ is the tensor operator
and we have pulled out a factor $N(0)=\frac{2 m^{\star}
k_{\text{F}}}{\pi^2}$, the density of states at the
Fermi surface, in order to make the Fermi liquid parameters $F_l,
F_l^\prime, G_l, G^\prime_l, H_l$ and $H_l^\prime$ dimensionless.
The effective mass of the quasiparticles is defined as $m^\star =
k_{\text{F}} / v_{\text{F}}$. We will discuss tensor interactions
in Section~6, but in order to simplify the discussion, we suppress
them in the derivation of the constraints. It is straightforward to
generalize the derivation and include them. Finally, since we
consider infinite nuclear matter, the spin-orbit interaction can be
neglected.

As in effective field theories, the Fermi liquid
parameters are determined by comparison with experiments. For
nuclear matter we have the following relations for the
incompressibility, the effective mass and the symmetry
energy~\cite{Landau,Migdal}:
\begin{align}
K &= \frac{3 \: \hbar^2 k_{\text{F}}^2}{m^\star} \: ( 1 + F_0 ), \\
\frac{m^\star}{m} &= 1 + F_1 / 3, \mbox{ and} \\
E_{\text{sym}} &= \frac{\hbar^2 k_{\text{F}}^2}{6 \: m^\star} \: ( 1 +
F_0^\prime ).
\end{align}
In order to establish the connection between the quasiparticle
interaction and the quasiparticle scattering amplitude, we consider
the leading particle-hole reducible contributions to the full
vertex function. We denote the bare particle-hole vertex
by $B(p,p^\prime;q)$, where the momenta $p, p^\prime$,
etc. and $q$ are 4-momenta, $p=(\varepsilon,{\mathbf p})$ and
$q=(\omega,{\mathbf q})$.
\begin{equation}
B(p,p^\prime;q) \quad =
\hspace{1cm} \parbox[c][4cm]{4cm}{\begin{fmfgraph*}(20,25)
\fmfpen{1.5}
\fmfleft{p1,pp1}
\fmflabel{$p+\frac{q}{2}$}{p1}
\fmflabel{$p^\prime+\frac{q}{2}$}{pp1}
\fmfright{p2,pp2}
\fmflabel{$p-\frac{q}{2}$}{p2}
\fmflabel{$p^\prime-\frac{q}{2}$}{pp2}
\fmf{fermion}{p1,v4}
\fmf{fermion}{v1,p2}
\fmf{fermion}{v3,pp1}
\fmf{fermion}{pp2,v2}
\fmf{plain}{s1,s2}
\fmfpoly{empty,smooth,tension=0.6}{v1,s1,v2,v3,s2,v4}
\end{fmfgraph*}}
\end{equation}
There are two possible ways to join two particle-hole vertices with
a particle-hole loop~\footnote{Note that in
Eq. (\ref{ZSprime}) we have used the antisymmetry
of the bare vertex $B(1+2,3+4;1-2)=-B(1+4,3+2,1-4)$.}:
\begin{multline}
\hspace{-0.5cm} -i \hspace{-0.1cm} \int \hspace{-0.1cm} \frac{d^4
p^{\prime\prime}}{(2 \pi)^4} \; B(p,p^{\prime\prime};q) \;
G(p^{\prime\prime}+\frac{q}{2}) \; G(p^{\prime\prime}-\frac{q}{2}) \;
B(p^{\prime\prime},p^\prime;q) = \hspace{0.5cm}
\parbox[c][6.25cm]{3.25cm}{\begin{fmfgraph*}(28.5,47.5)
\fmfpen{1.5}
\fmfleft{p1,pp1}
\fmflabel{$p+\frac{q}{2}$}{p1}
\fmflabel{$p^\prime+\frac{q}{2}$}{pp1}
\fmfright{p2,pp2}
\fmflabel{$p-\frac{q}{2}$}{p2}
\fmflabel{$p^\prime-\frac{q}{2}$}{pp2}
\fmf{fermion}{p1,v4}
\fmf{fermion}{v1,p2}
\fmf{fermion}{u3,pp1}
\fmf{fermion}{pp2,u2}
\fmf{fermion,left=0.5,tension=0.5,label=$p^{\prime\prime}-\frac{q}{2}$,label.side=left}{u1,v2}
\fmf{fermion,left=0.5,tension=0.5,label=$p^{\prime\prime}+\frac{q}{2}$,label.side=left}{v3,u4}
\fmf{plain}{s1,s2}
\fmf{plain}{s3,s4}
\fmfpoly{empty,smooth,tension=0.6}{v1,s1,v2,v3,s2,v4}
\fmfpoly{empty,smooth,tension=0.6}{u1,s3,u2,u3,s4,u4}
\end{fmfgraph*}}
\label{ZS}
\end{multline}
\begin{multline}
-i \int \frac{d^4 p^{\prime\prime}}{(2 \pi)^4} \;
B(\frac{p+p^\prime+q}{2},p^{\prime\prime};p-p^\prime) \;
G(p^{\prime\prime}+\frac{p-p^\prime}{2}) \;
G(p^{\prime\prime}-\frac{p-p^\prime}{2}) \\
\times B(p^{\prime\prime},\frac{p+p^\prime-q}{2};p-p^\prime) \quad =
\hspace{0.8cm} \parbox[c][4.5cm]{5.5cm}{\begin{fmfgraph*}(50,30)
\fmfpen{1.5}
\fmfleft{p1,pp1}
\fmflabel{$p+\frac{q}{2}$}{p1}
\fmflabel{$p^\prime+\frac{q}{2}$}{pp1}
\fmfright{p2,pp2}
\fmflabel{$p-\frac{q}{2}$}{p2}
\fmflabel{$p^\prime-\frac{q}{2}$}{pp2}
\fmf{fermion}{p1,v4}
\fmf{fermion}{u1,p2}
\fmf{fermion}{v3,pp1}
\fmf{fermion}{pp2,u2}
\fmf{fermion,right=0.5,tension=0.5,label=$p^{\prime\prime}-\frac{p-p^\prime}{2}$,label.side=right}{u3,v2}
\fmf{fermion,right=0.5,tension=0.5,label=$p^{\prime\prime}+\frac{p-p^\prime}{2}$,label.side=right}{v1,u4}
\fmf{plain}{s1,s2}
\fmf{plain}{s3,s4}
\fmfpoly{empty,smooth,tension=0.6}{v1,s1,v2,v3,s2,v4}
\fmfpoly{empty,smooth,tension=0.6}{u1,s3,u2,u3,s4,u4}
\end{fmfgraph*}} \quad .
\label{ZSprime}
\end{multline}
In the recent literature, the first channel, Eq. (\ref{ZS}), is
referred to as the zero sound channel (ZS), while the second
one is called ZS$^\prime$. The ZS$^\prime$ diagram,
Eq. (\ref{ZSprime}), is the exchange diagram to the ZS graph.
Landau wrote down a Bethe-Salpeter equation, which sums
the particle-hole ladders in the ZS channel. This equation relates
the full particle-hole vertex $\Gamma(p,p^\prime;q)$ to the ZS
particle-hole irreducible one $\tilde{\Gamma}(p,p^\prime;q)$:
\begin{equation}
\hspace{0.7cm} \parbox[c][4cm]{4cm}{\begin{fmfgraph*}(20,25)
\fmfpen{1.5}
\fmfleft{p1,pp1}
\fmflabel{$p+\frac{q}{2}$}{p1}
\fmflabel{$p^\prime+\frac{q}{2}$}{pp1}
\fmfright{p2,pp2}
\fmflabel{$p-\frac{q}{2}$}{p2}
\fmflabel{$p^\prime-\frac{q}{2}$}{pp2}
\fmf{fermion}{p1,v4}
\fmf{fermion}{v1,p2}
\fmf{fermion}{v3,pp1}
\fmf{fermion}{pp2,v2}
\fmf{phantom}{s1,s2}
\fmfpoly{smooth,tension=0.6}{v1,s1,v2,v3,s2,v4}
\fmfv{label=\Huge$\Gamma$,label.dist=-0.75cm}{s2}
\end{fmfgraph*}} \hspace{-1.1cm} = \hspace{0.9cm}
\parbox[c][4cm]{4cm}{\begin{fmfgraph*}(20,25)
\fmfpen{1.5}
\fmfleft{p1,pp1}
\fmflabel{$p+\frac{q}{2}$}{p1}
\fmflabel{$p^\prime+\frac{q}{2}$}{pp1}
\fmfright{p2,pp2}
\fmflabel{$p-\frac{q}{2}$}{p2}
\fmflabel{$p^\prime-\frac{q}{2}$}{pp2}
\fmf{fermion}{p1,v4}
\fmf{fermion}{v1,p2}
\fmf{fermion}{v3,pp1}
\fmf{fermion}{pp2,v2}
\fmf{phantom}{s1,s2}
\fmfpoly{smooth,tension=0.6}{v1,s1,v2,v3,s2,v4}
\fmfv{label=\Huge$\tilde{\Gamma}$,label.dist=-0.775cm}{s2}
\end{fmfgraph*}} \hspace{-1.1cm} + \hspace{0.9cm}
\parbox[c][6cm]{3.5cm}{\begin{fmfgraph*}(30,50)
\fmfpen{1.5}
\fmfleft{p1,pp1}
\fmflabel{$p+\frac{q}{2}$}{p1}
\fmflabel{$p^\prime+\frac{q}{2}$}{pp1}
\fmfright{p2,pp2}
\fmflabel{$p-\frac{q}{2}$}{p2}
\fmflabel{$p^\prime-\frac{q}{2}$}{pp2}
\fmf{fermion}{p1,v4}
\fmf{fermion}{v1,p2}
\fmf{fermion}{u3,pp1}
\fmf{fermion}{pp2,u2}
\fmf{fermion,left=0.5,tension=0.5,label=$p^{\prime\prime}-\frac{q}{2}$,label.side=left}{u1,v2}
\fmf{fermion,left=0.5,tension=0.5,label=$p^{\prime\prime}+\frac{q}{2}$,label.side=left}{v3,u4}
\fmf{phantom}{s1,s2}
\fmf{phantom}{s3,s4}
\fmfpoly{smooth,tension=0.6}{v1,s1,v2,v3,s2,v4}
\fmfpoly{smooth,tension=0.6}{u1,s3,u2,u3,s4,u4}
\fmfv{label=\Huge$\tilde{\Gamma}$,label.dist=0.25cm,label.angle=0}{s2}
\fmfv{label=\Huge$\Gamma$,label.dist=0.25cm,label.angle=-4}{s4}
\end{fmfgraph*}} \quad .
\label{SAinGamma}
\end{equation}
The Bethe-Salpeter equation in the ZS channel reads
\begin{multline}
\Gamma(p,p^\prime;q) = \tilde{\Gamma}(p,p^\prime;q) -i \: \int \: \frac{d^4
p^{\prime\prime}}{(2 \pi)^4} \: \tilde{\Gamma}(p,p^{\prime\prime};q) \\
\times G(p^{\prime\prime}+\frac{q}{2}) \: G(p^{\prime\prime}-\frac{q}{2}) \:
\Gamma(p^{\prime\prime},p^\prime;q) .
\end{multline}
As argued above, we set $p$ and $p^\prime$ on the Fermi surface and
let $q \to 0$. In finite nuclei, typical momentum transfers
$|{\mathbf q}|$ are of the order of the inverse size of the
nucleus. Therefore, on physical grounds, $|{\mathbf q}|
\sim 1/R \sim A^{-1/3}$ vanishes in nuclear matter~\cite{Migdal}.
Landau noticed that the product of propagators
$G(p^{\prime\prime}+\frac{q}{2}) \;
G(p^{\prime\prime}-\frac{q}{2})$ is singular in the limit
$|{\mathbf q}|\to 0$ and $\omega \to 0$ (see e.g.~\cite{AGD}) and
therefore $\tilde{\Gamma}$ is by construction finite as $q \to 0$.
The singularity is due to the quasiparticle poles in the
propagators:
\begin{align}
& G(p^{\prime\prime}+\frac{q}{2}) \; G(p^{\prime\prime}-\frac{q}{2})
= \: \frac{z}{\varepsilon^{\prime\prime} +
\omega/2 - v_{\text{F}} (|{\mathbf p^{\prime\prime}}+{\mathbf q}/2|
- k_{\text{F}}) + i \delta_{p^{\prime\prime}+\frac{q}{2}}}
\nonumber \\
& \: \times \: \frac{z}{\varepsilon^{\prime\prime} -
\omega/2 - v_{\text{F}} (|{\mathbf p^{\prime\prime}}-{\mathbf q}/2| -
k_{\text{F}}) + i \delta_{p^{\prime\prime}-\frac{q}{2}}} + \:
\mbox{multi-pair background} \nonumber \\[1mm]
& \: \: = \: \frac{2 \pi i z^2}{v_{\text{F}}} \frac{v_{\text{F}} \:
\hat{{\mathbf p}}^{\prime\prime} \cdot {\mathbf q}}{\omega - v_{\text{F}} \:
\hat{{\mathbf p}}^{\prime\prime} \cdot {\mathbf q}} \: \delta
(\varepsilon^{\prime\prime}) \: \delta (|{\mathbf
p^{\prime\prime}}| - k_{\text{F}}) \: + \: \mbox{non-singular} \:
\phi(p^{\prime\prime}),
\label{PP}
\end{align}
where the quasiparticle energy
is measured relative to the Fermi energy $\mu$. We note that the
singular part, which is due to the quasiparticle piece of the Green
functions, vanishes in the limit $|{\mathbf q}| \to 0$ and $\omega\to
0$ with $|{\mathbf q}| / \omega \to 0$.
Therefore, one can eliminate all quasiparticle-quasihole reducible
contributions in a given channel by taking this limit.

The singularity of the ZS particle-hole propagator is reflected in
the dependence of the coefficient of the delta functions in
Eq. (\ref{PP}) on the order of the
limits $|{\mathbf q}| \to 0$ and $\omega \to 0$.
The $|{\mathbf q}|$ and $\omega$ limits of
the particle-hole vertex are defined as:
\begin{align}
\Gamma^\omega (p,p^\prime) &= \lim_{\omega \to 0} \: ( \: \Gamma
(p,p^\prime;q) \: | _{|{\mathbf q}|=0} \: ), \mbox{ and} \\
\Gamma^q (p,p^\prime) &= \lim_{|{\mathbf q}| \to 0} \: ( \: \Gamma
(p,p^\prime;q) \: | _{\omega=0} \: ).
\end{align}
In the $\omega$ limit the singular part in Eq.
(\ref{PP}) vanishes. Thus, from Eq. (\ref{SAinGamma}) it follows that
$\Gamma^\omega$ itself is obtained by solving a Bethe-Salpeter
equation, which sums the ZS particle-hole ladders with the non-singular
part $\phi$ only. Consequently, $\Gamma^\omega$
is {\it quasi}particle-{\it quasi}hole irreducible in the ZS channel.

With this at hand, one can eliminate $\tilde{\Gamma}$ and the
non-singular $\phi$ to obtain the following quasiparticle-quasihole
analogue of Eq. (\ref{SAinGamma}) for $\Gamma$, at
$T=0$~\cite{AGD}:
\begin{eqnarray}
\hspace{-0.8cm} \Gamma_{{\bm \sigma} \cdot {\bm \sigma^{\prime}},
{\bm \tau} \cdot {\bm \tau^{\prime}}}(p,p^\prime;q) =
\Gamma^\omega_{{\bm \sigma} \cdot {\bm \sigma^{\prime}},
{\bm \tau} \cdot {\bm \tau^{\prime}}}(p,p^\prime) &+& N(0) \: z^2
\frac{1}{4} \: \text{Tr}_{{\bm \sigma^{\prime\prime}} {\bm
\tau^{\prime\prime}}} \int \frac{d \Omega_{{\mathbf
p^{\prime\prime}}}}{4 \pi} \: \Gamma^\omega_{{\bm \sigma} \cdot
{\bm \sigma^{\prime\prime}},
{\bm \tau} \cdot {\bm \tau^{\prime\prime}}}(p,p^{\prime\prime})
\nonumber \\[2mm]
&& \frac{v_{\text{F}} \: \hat{{\mathbf p}}^{\prime\prime} \cdot {\mathbf q}}
{\omega - v_{\text{F}} \: \hat{{\mathbf p}}^{\prime\prime}
\cdot {\mathbf q}} \:
\Gamma_{{\bm \sigma^{\prime\prime}} \cdot {\bm \sigma^{\prime}},
{\bm \tau^{\prime\prime}} \cdot{\bm \tau^{\prime}}}
(p^{\prime\prime},p^\prime;q).
\label{SA}
\end{eqnarray}
Diagrammatically this equation corresponds to
\begin{equation}
\hspace{0.75cm} \parbox[c][4cm]{4cm}{\begin{fmfgraph*}(20,25)
\fmfpen{1.5}
\fmfleft{p1,pp1}
\fmflabel{$p+\frac{q}{2}$}{p1}
\fmflabel{$p^\prime+\frac{q}{2}$}{pp1}
\fmfright{p2,pp2}
\fmflabel{$p-\frac{q}{2}$}{p2}
\fmflabel{$p^\prime-\frac{q}{2}$}{pp2}
\fmf{fermion}{p1,v4}
\fmf{fermion}{v1,p2}
\fmf{fermion}{v3,pp1}
\fmf{fermion}{pp2,v2}
\fmf{phantom}{s1,s2}
\fmfpoly{smooth,tension=0.6}{v1,s1,v2,v3,s2,v4}
\fmfv{label=\Huge$\Gamma$,label.dist=-0.75cm}{s2}
\end{fmfgraph*}} \hspace{-1.1cm} = \hspace{0.9cm}
\parbox[c][4cm]{4cm}{\begin{fmfgraph*}(20,25)
\fmfpen{1.5}
\fmfleft{p1,pp1}
\fmflabel{$p+\frac{q}{2}$}{p1}
\fmflabel{$p^\prime+\frac{q}{2}$}{pp1}
\fmfright{p2,pp2}
\fmflabel{$p-\frac{q}{2}$}{p2}
\fmflabel{$p^\prime-\frac{q}{2}$}{pp2}
\fmf{fermion}{p1,v4}
\fmf{fermion}{v1,p2}
\fmf{fermion}{v3,pp1}
\fmf{fermion}{pp2,v2}
\fmf{plain}{s1,s2}
\fmfpoly{smooth,tension=0.6}{v1,s1,v2,v3,s2,v4}
\end{fmfgraph*}} \hspace{-1.1cm} + \hspace{0.9cm}
\parbox[c][6.25cm]{3.5cm}{\begin{fmfgraph*}(30,50)
\fmfpen{1.5}
\fmfleft{p1,pp1}
\fmflabel{$p+\frac{q}{2}$}{p1}
\fmflabel{$p^\prime+\frac{q}{2}$}{pp1}
\fmfright{p2,pp2}
\fmflabel{$p-\frac{q}{2}$}{p2}
\fmflabel{$p^\prime-\frac{q}{2}$}{pp2}
\fmf{fermion}{p1,v4}
\fmf{fermion}{v1,p2}
\fmf{fermion}{u3,pp1}
\fmf{fermion}{pp2,u2}
\fmf{fermion,left=0.5,tension=0.5,label=$p^{\prime\prime}-\frac{q}{2}$,label.side=left}{u1,v2}
\fmf{fermion,left=0.5,tension=0.5,label=$p^{\prime\prime}+\frac{q}{2}$,label.side=left}{v3,u4}
\fmf{plain}{s1,s2}
\fmf{phantom}{s3,s4}
\fmfpoly{smooth,tension=0.6}{v1,s1,v2,v3,s2,v4}
\fmfpoly{smooth,tension=0.6}{u1,s3,u2,u3,s4,u4}
\fmfv{label=\Huge$\Gamma$,label.dist=0.25cm,label.angle=-4}{s4}
\end{fmfgraph*}} \quad ,
\label{SAinF}
\end{equation}
where $\Gamma^\omega$ is denoted by a blob with a line across. The
line is drawn perpendicular to the channel, in which
$\Gamma^\omega$ is quasiparticle-quasihole irreducible~\footnote{We
use the notation that the particle-hole propagators in diagrams
with the crossed blob correspond to the singular
quasiparticle-quasihole part only.}.

The $|{\mathbf q}|$ limit $\Gamma^q$ corresponds to the full
particle-hole vertex for $q=0$, i.e. scattering of
quasiparticles strictly on the Fermi surface with vanishing momentum
transfer $|{\mathbf q}| \to 0$.
Thus, Eq. (\ref{SA}) can be used to relate the two limits:
\begin{equation}
\Gamma^q_{{\bm \sigma} \cdot {\bm \sigma^{\prime}}, {\bm \tau}
\cdot {\bm \tau^{\prime}}} = \Gamma^\omega_{{\bm \sigma} \cdot {\bm
\sigma^{\prime}}, {\bm \tau} \cdot {\bm \tau^{\prime}}} -
N(0) \, z^2 \: \frac{1}{4}\int\frac{d\Omega_{p^{\prime\prime}}}{4\pi} \:
\text{Tr}_{{\bm
\sigma^{\prime\prime}} {\bm \tau^{\prime\prime}}} \: \Gamma^\omega_{{\bm
\sigma} \cdot {\bm \sigma^{\prime\prime}}, {\bm \tau} \cdot {\bm
\tau^{\prime\prime}}} \: \Gamma^q_{{\bm
\sigma^{\prime\prime}} \cdot {\bm \sigma^{\prime}}, {\bm
\tau^{\prime\prime}} \cdot {\bm \tau^{\prime}}} \, .
\label{LL}
\end{equation}
The quasiparticle-quasihole irreducible vertex can be identified
with the quasiparticle interaction introduced above~\cite{AGD},
$N(0) \, z^2 \: \Gamma^\omega (p,p^\prime) = {\mathcal F} (\theta)$, while
$N(0) \, z^2 \:\Gamma^q (p,p^\prime) = {\mathcal A} (\theta)$
is the quasiparticle forward scattering amplitude. The factor $z$
is the spectral strength at the quasiparticle pole. By inserting
this into Eq. (\ref{LL}) and expanding the angular dependence of
${\mathcal F}(\theta)$ and ${\mathcal A}(\theta)$ on Legendre
polynomials, we arrive at a set of algebraic equations for the
scattering amplitude with the solution:
\begin{multline}
{\mathcal A} (\theta) = \sum_{l} \biggl( \frac{F_l}{1 + F_l / (2 l +
1)} + \frac{F_l^{\prime}}{1 + F_l^{\prime} / (2 l + 1)} \: {\bm \tau}
\cdot {\bm \tau^{\prime}} \\ + \frac{G_l}{1 + G_l / (2 l + 1)} \:
{\bm \sigma} \cdot {\bm \sigma^{\prime}} + \frac{G_l^{\prime}}{1 +
G_l^{\prime} / (2 l + 1)} \: {\bm \tau} \cdot {\bm \tau^{\prime}} \:
{\bm \sigma} \cdot {\bm \sigma^{\prime}} \biggr) P_l (\cos \theta).
\end{multline}
The antisymmetry of the quasiparticle scattering amplitude implies
two Pauli principle sum rules~\cite{Landau,BD} for the Fermi liquid
parameters, corresponding to scattering at vanishing relative momentum
in singlet-odd and triplet-odd states:
\begin{align}
& \sum_{l} \biggl( \frac{F_l}{1 + F_l/(2 l + 1)} + \frac{F_l^{\prime}}{1
+ F_l^{\prime} / (2 l + 1)} \nonumber \\
& \hspace{4cm} + \frac{G_l}{1 + G_l / (2 l + 1)} +
\frac{G_l^{\prime}}{1 + G_l^{\prime} / (2 l + 1)} \biggr) = 0
\label{SR1} \\[2mm]
& \sum_{l} \biggl( \frac{F_l}{1 + F_l / (2 l + 1)} - 3 \:
\frac{F_l^{\prime}}{1 + F_l^{\prime} / (2 l + 1)} \nonumber \\
& \hspace{3.6cm} - 3 \: \frac{G_l}{1 + G_l / (2 l + 1)} + 9 \:
\frac{G_l^{\prime}}{1 + G_l^{\prime} / (2 l + 1)} \biggr) = 0.
\label{SR2}
\end{align}
It is important to note that the quasiparticle interaction is
strictly speaking defined only in the Landau limit $q = 0$.
This is reflected in
the one pion exchange (OPE) contribution (direct and exchange) to
$\Gamma^\omega$, where the direct tensor interaction, which is proportional to
${\mathbf q}^2$, vanishes in the Landau limit. For later use we give
the one pion exchange contribution to $\Gamma^\omega$:
\begin{multline}
\Gamma^{\text{OPE}}_{{\bm \sigma} \cdot {\bm \sigma^{\prime}}, {\bm \tau}
\cdot {\bm \tau^{\prime}}} (p,p^\prime;q) = - \frac{f^2}{3 \:
m_\pi^2} \: {\bm \tau} \cdot {\bm \tau^{\prime}} \biggl\{ {\mathbf q}^2
\frac{S_{12} ({\mathbf q})}{{\mathbf q^2} +m^2_\pi} -
\frac{m^2_\pi \: {\bm \sigma}
\cdot {\bm \sigma^{\prime}}}{{\mathbf q^2} +m^2_\pi} \biggr\} \\[2mm]
+ \frac{f^2}{3 \:
m_\pi^2} \: \frac{3 - {\bm \tau} \cdot {\bm \tau^{\prime}}}{2}
\biggl\{ ({\mathbf p} -{\mathbf p^\prime})^2
\frac{S_{12} ({\mathbf p} - {\mathbf p^\prime})}{({\mathbf
p} -{\mathbf p^\prime})^2 +m^2_\pi} -\frac{1}{2} \: \frac{m^2_\pi
\: ( 3 - {\bm \sigma} \cdot {\bm \sigma^{\prime}})}{({\mathbf p}
-{\mathbf p^\prime})^2 +m^2_\pi} \biggr\} \,.
\label{OPE}
\end{multline}

\section{The Induced Interaction}

The quasiparticle scattering amplitude includes particle-hole diagrams
in the ZS channel to all orders. Therefore, if one were to use a
finite set of diagrams for the quasiparticle-quasihole irreducible
vertex $\Gamma^\omega$, e.g. the Hartree-Fock approximation,
Eq. (\ref{HF}), then the corresponding quasiparticle
scattering amplitude, obtained by solving Eq. (\ref{SA}), would not obey the
Pauli principle. This is because the particle-hole diagrams in the
ZS$^\prime$ channel, which, as discussed above are the exchange
diagrams to those in the ZS channel, are not iterated.
Thus, in order to obey the Pauli principle, it
is necessary to iterate the ZS$^\prime$ channel to all orders as well.
This is done by the induced interaction, which was invented by Babu and
Brown~\cite{BB} and applied to nuclear matter by
Sj\"oberg~\cite{BBfornuclmat}. Here we give a diagrammatically
motivated heuristic derivation of the induced interaction.

The integral equation for $\Gamma^\omega$ must be constructed in
such a way that it generates all possible ZS and ZS$^\prime$ joined
diagrams for the quasiparticle scattering amplitude.
To third order these are:
\begin{align}
& \parbox[c]{1.75cm}{\begin{fmfgraph}(15,15)
\fmfpen{1.5}
\fmfleft{p1,pp1}
\fmfright{p2,pp2}
\fmf{plain}{p1,v1}
\fmf{plain}{v1,p2}
\fmf{plain}{v1,pp1}
\fmf{plain}{pp2,v1}
\fmfdotn{v}{1}
\end{fmfgraph}}
\parbox[c]{2cm}{\begin{fmfgraph}(20,30)
\fmfpen{1.5}
\fmfleft{p1,pp1}
\fmfright{p2,pp2}
\fmf{plain}{p1,v1}
\fmf{plain}{v1,p2}
\fmf{plain}{v2,pp1}
\fmf{plain}{pp2,v2}
\fmf{plain,left=0.5,tension=0.5,width=3.5}{v1,v2}
\fmf{plain,left=0.5,tension=0.5,width=3.5}{v2,v1}
\fmfdotn{v}{2}
\end{fmfgraph}}
\parbox[c]{3cm}{\begin{fmfgraph}(30,20)
\fmfpen{1.5}
\fmfleft{p1,pp1}
\fmfright{p2,pp2}
\fmf{plain}{p1,v1}
\fmf{plain}{v2,p2}
\fmf{plain}{v1,pp1}
\fmf{plain}{pp2,v2}
\fmf{plain,right=0.5,tension=0.5}{v2,v1}
\fmf{plain,right=0.5,tension=0.5}{v1,v2}
\fmfforce{(0.25w,0.5h)}{v1}
\fmfforce{(0.75w,0.5h)}{v2}
\fmfdotn{v}{2}
\end{fmfgraph}}
\parbox[c]{3cm}{\begin{fmfgraph}(30,30)
\fmfpen{1.5}
\fmfleft{p1,pp1}
\fmfright{p2,pp2}
\fmf{plain}{p1,v1}
\fmf{plain}{v3,p2}
\fmf{plain}{v2,pp1}
\fmf{plain}{pp2,v3}
\fmf{plain,left=0.5,tension=0.5}{v1,v2}
\fmf{plain,left=0.5,tension=0.5}{v2,v1}
\fmf{plain,left=0.2,tension=0.2}{v3,v1}
\fmf{plain,left=0.2,tension=0.2}{v2,v3}
\fmfforce{(0.2w,0.25h)}{v1}
\fmfforce{(0.2w,0.75h)}{v2}
\fmfforce{(0.8w,0.5h)}{v3}
\fmfdotn{v}{3}
\end{fmfgraph}}
\parbox[c]{3cm}{\begin{fmfgraph}(30,30)
\fmfpen{1.5}
\fmfleft{p1,pp1}
\fmfright{p2,pp2}
\fmf{plain}{p1,v3}
\fmf{plain}{v1,p2}
\fmf{plain}{v3,pp1}
\fmf{plain}{pp2,v2}
\fmf{plain,left=0.5,tension=0.5}{v1,v2}
\fmf{plain,left=0.5,tension=0.5}{v2,v1}
\fmf{plain,left=0.2,tension=0.2}{v1,v3}
\fmf{plain,left=0.2,tension=0.2}{v3,v2}
\fmfdotn{v}{3}
\fmfforce{(0.8w,0.25h)}{v1}
\fmfforce{(0.8w,0.75h)}{v2}
\fmfforce{(0.2w,0.5h)}{v3}
\end{fmfgraph}} \nonumber \\
& \hspace{0.5cm} (a) \hspace{1.45cm} (b) \hspace{2cm} (c)
\hspace{2.45cm} (d) \hspace{2.5cm} (e) \nonumber \\[1.5mm]
& \parbox[c]{3cm}{\begin{fmfgraph}(30,30)
\fmfpen{1.5}
\fmfleft{p1,pp1}
\fmfright{p2,pp2}
\fmf{plain}{p1,v3}
\fmf{plain}{v3,p2}
\fmf{plain}{v1,pp1}
\fmf{plain}{pp2,v2}
\fmf{plain,right=0.5,tension=0.5}{v2,v1}
\fmf{plain,right=0.5,tension=0.5}{v1,v2}
\fmf{plain,left=0.2,tension=0.2,width=3.5}{v3,v1}
\fmf{plain,left=0.2,tension=0.2,width=3.5}{v2,v3}
\fmfforce{(0.25w,0.8h)}{v1}
\fmfforce{(0.75w,0.8h)}{v2}
\fmfforce{(0.5w,0.2h)}{v3}
\fmfdotn{v}{3}
\end{fmfgraph}}
\parbox[c]{3cm}{\begin{fmfgraph}(30,30)
\fmfpen{1.5}
\fmfleft{p1,pp1}
\fmfright{p2,pp2}
\fmf{plain}{p1,v1}
\fmf{plain}{v2,p2}
\fmf{plain}{v3,pp1}
\fmf{plain}{pp2,v3}
\fmf{plain,right=0.5,tension=0.5}{v2,v1}
\fmf{plain,right=0.5,tension=0.5}{v1,v2}
\fmf{plain,left=0.2,tension=0.2,width=3.5}{v1,v3}
\fmf{plain,left=0.2,tension=0.2,width=3.5}{v3,v2}
\fmfforce{(0.27w,0.2h)}{v1}
\fmfforce{(0.75w,0.2h)}{v2}
\fmfforce{(0.5w,0.8h)}{v3}
\fmfdotn{v}{3}
\end{fmfgraph}}
\parbox[c]{2.25cm}{\begin{fmfgraph}(25,45)
\fmfpen{1.5}
\fmfleft{p1,pp1}
\fmfright{p2,pp2}
\fmf{plain}{p1,v1}
\fmf{plain}{v1,p2}
\fmf{plain}{v3,pp1}
\fmf{plain}{pp2,v3}
\fmf{plain,left=0.5,tension=0.5,width=3.5}{v1,v2}
\fmf{plain,left=0.5,tension=0.5,width=3.5}{v2,v1}
\fmf{plain,left=0.5,tension=0.5,width=3.5}{v2,v3}
\fmf{plain,left=0.5,tension=0.5,width=3.5}{v3,v2}
\fmfforce{(0.5w,0.15h)}{v1}
\fmfforce{(0.5w,0.5h)}{v2}
\fmfforce{(0.5w,0.85h)}{v3}
\fmfdotn{v}{3}
\end{fmfgraph}}
\parbox[c]{5cm}{\begin{fmfgraph}(50,25)
\fmfpen{1.5}
\fmfleft{p1,pp1}
\fmfright{p2,pp2}
\fmf{plain}{p1,v1}
\fmf{plain}{v3,p2}
\fmf{plain}{v1,pp1}
\fmf{plain}{pp2,v3}
\fmf{plain,right=0.5,tension=0.5}{v2,v1}
\fmf{plain,right=0.5,tension=0.5}{v1,v2}
\fmf{plain,right=0.5,tension=0.5}{v3,v2}
\fmf{plain,right=0.5,tension=0.5}{v2,v3}
\fmfforce{(0.2w,0.5h)}{v1}
\fmfforce{(0.5w,0.5h)}{v2}
\fmfforce{(0.8w,0.5h)}{v3}
\fmfdotn{v}{3}
\end{fmfgraph}} \nonumber \\
& \hspace{1.25cm} (f) \hspace{2.4cm} (g) \hspace{2.25cm} (h) \hspace{3cm} (i)
\label{parquet}
\end{align}
Using an antisymmetric, particle-hole irreducible vertex function
in Eq. (\ref{parquet}) guarantees the antisymmetry of the quasiparticle
scattering amplitude. We have marked the propagators
$G(p^{\prime\prime}+\frac{q}{2}) \;
G(p^{\prime\prime}-\frac{q}{2})$, that are generated 
by solving the Bethe-Salpeter equation, Eq. (\ref{SAinGamma}), with
thick lines. The diagrams with only thin lines are
contained in $\tilde{\Gamma}$. All of these can be constructed from
a ZS$^\prime$ ladder sum, where the vertices are lower order
diagrams of $\tilde{\Gamma}$ rotated by 90 degrees. To second order
$\tilde{\Gamma}$ consists of the one ZS$^\prime$ bubble only,
diagram (c), to third order $\tilde{\Gamma}$ also includes the two
ZS$^\prime$ bubble string, diagram (i), and the diagrams (d) and
(e). The latter are constructed by taking a second (lower) order
diagram, the one ZS$^\prime$ bubble, diagram (c), rotating it by 90
degrees, and then inserting it as left or right vertex into the one
ZS$^\prime$ bubble. Thus, the integral equation for
$\tilde{\Gamma}$, for a system with spin only, reads:
\begin{align}
&\tilde{\Gamma}_{{\bm \sigma} \cdot {\bm
\sigma^{\prime}}}(p,p^\prime;q) =
I_{{\bm \sigma} \cdot {\bm \sigma^{\prime}}}(p,p^\prime;q) -
\frac{1}{2} \: (1+{\bm \sigma} \cdot {\bm \sigma^{\prime}})
\nonumber \\[1mm]
& \: \times \biggl\{
\frac{1}{2} \: \text{Tr}_{{\bm \sigma^{\prime\prime}}}
\int \frac{-i \, d^4 p^{\prime\prime}}{(2 \pi)^4} \,
\tilde{\Gamma}_{{\bm \sigma} \cdot {\bm \sigma^{\prime\prime}}}
(\frac{p+p^\prime+q}{2},p^{\prime\prime};p-p^\prime) \,
G(p^{\prime\prime}+\frac{p-p^\prime}{2}) \nonumber \\[1mm]
& \: \: \: \times G(p^{\prime\prime}-\frac{p-p^\prime}{2}) \,
\tilde{\Gamma}_{{\bm \sigma^{\prime\prime}} \cdot {\bm
\sigma^{\prime}}}(p^{\prime\prime},\frac{p+p^\prime-q}{2};p-p^\prime)
+ \tilde{\Gamma} \, G^2 \, \tilde{\Gamma} \, G^2 \, \tilde{\Gamma} +
\ldots \biggr\} \, ,
\label{Gammatilde}
\end{align}
where we have denoted the antisymmetric, ZS and ZS$^\prime$
particle-hole irreducible vertex with $I$. The spin operator ${\bm
\sigma}$ in the brackets of Eq. (\ref{Gammatilde}) is contracted
with the spinors of the left particle-hole pair with momenta
$p+q/2$ and $p^\prime+q/2$, whereas ${\bm \sigma}$ in the left hand
side and in $I$ is contracted with the bottom particle-hole pair
spinors with momenta $p \pm q/2$. The recoupling
between the two particle-hole channels is accounted for by
including the spin exchange operator $P_{{\bm
\sigma}} = 1/2 \: (1+{\bm
\sigma} \cdot {\bm \sigma^{\prime}})$. Since $\tilde{\Gamma}$ is
finite, we can take the limit $q \to 0$ in Eq. (\ref{Gammatilde})
and obtain for ${\mathbf p} \approx {\mathbf
p^\prime}$~\footnote{To guarantee continuity in the forward
scattering amplitude the limit $q \to 0$ has to be performed before
taking ${\mathbf p} \to {\mathbf p^\prime}$~\cite{Mermin}.}
\begin{align}
&\tilde{\Gamma}_{{\bm \sigma} \cdot {\bm
\sigma^{\prime}}}(p,p^\prime) =
I_{{\bm \sigma} \cdot {\bm \sigma^{\prime}}}(p,p^\prime) -
\frac{1}{2} \: (1+{\bm \sigma} \cdot {\bm \sigma^{\prime}}) \nonumber \\[1mm]
& \: \times \biggl\{
\frac{1}{2} \: \text{Tr}_{{\bm \sigma^{\prime\prime}}}
\biggl( N(0) \: z^2 \: \int \frac{d \Omega_{{\mathbf
p^{\prime\prime}}}}{4 \pi} \, \tilde{\Gamma}_{{\bm \sigma} \cdot {\bm
\sigma^{\prime\prime}}} (\frac{p+p^\prime}{2},p^{\prime\prime};p-p^\prime)
\nonumber \\
& \hspace{2.5cm}
\times \frac{v_{\text{F}} \: \hat{{\mathbf p^{\prime\prime}}}
\cdot ({\mathbf
p} - {\mathbf p^\prime})}{\varepsilon - \varepsilon^\prime - v_{\text{F}} \:
\hat{{\mathbf p^{\prime\prime}}} \cdot ({\mathbf p} - {\mathbf
p^\prime})} \: \tilde{\Gamma}_{{\bm \sigma^{\prime\prime}} \cdot
{\bm \sigma^{\prime}}} (\frac{p+p^\prime}{2};p-p^\prime) \nonumber \\[1mm]
& \quad \: + \int \frac{-i
\, d^4 p^{\prime\prime}}{(2 \pi)^4} \,
\tilde{\Gamma}_{{\bm \sigma} \cdot {\bm \sigma^{\prime\prime}}}
(\frac{p+p^\prime}{2},p^{\prime\prime};p-p^\prime) \,
\phi(p^{\prime\prime}) \, \tilde{\Gamma}_{{\bm
\sigma^{\prime\prime}} \cdot {\bm
\sigma^{\prime}}}(p^{\prime\prime},\frac{p+p^\prime}{2};p-p^\prime)
\biggr) \nonumber \\
& \quad \: \:
+ \tilde{\Gamma} \: \bigl( ({\mathcal G} {\mathcal
G})_{\text{ZS}^\prime} + \phi \bigr) \: \tilde{\Gamma} \:
\bigl( ({\mathcal G} {\mathcal G})_{\text{ZS}^\prime} + \phi \bigr) \:
\tilde{\Gamma} + \ldots \biggr\} \, ,
\label{Gammatilde2}
\end{align}
where $({\mathcal G} {\mathcal G})_{\text{ZS}^\prime}$ denotes the
quasiparticle-quasihole part of the propagators in the ZS$^\prime$
channel. To both sides of Eq. (\ref{Gammatilde2}) we add the series
$\tilde{\Gamma} \, \phi \, \tilde{\Gamma} + \tilde{\Gamma} \, \phi
\,
\tilde{\Gamma} \, \phi \, \tilde{\Gamma} + \ldots$ and obtain
\begin{multline}
\Gamma^\omega = I + \tilde{\Gamma} \, \phi \,
\tilde{\Gamma} + \tilde{\Gamma} \, \phi \, \tilde{\Gamma} \, \phi \,
\tilde{\Gamma} + \ldots - \frac{1}{2} \: (1+{\bm \sigma} \cdot {\bm
\sigma^{\prime}}) \: \bigl\{ \, \tilde{\Gamma} \: \bigl( ({\mathcal G}
{\mathcal G})_{\text{ZS}^\prime} + \phi \bigr) \: \tilde{\Gamma} \\[1mm]
+ \tilde{\Gamma} \: \bigl( ({\mathcal G} {\mathcal
G})_{\text{ZS}^\prime} + \phi \bigr) \:
\tilde{\Gamma} \: \bigl( ({\mathcal G} {\mathcal
G})_{\text{ZS}^\prime} + \phi \bigr) \: \tilde{\Gamma} +
\ldots \bigr\} \, .
\end{multline}
By regrouping the terms we find
\begin{multline}
\Gamma^\omega = I + \bigl( 1 - \frac{1}{2} \: (1+{\bm \sigma} \cdot {\bm
\sigma^{\prime}}) \bigr) \: \tilde{\Gamma} \, \phi \,
\: \frac{1}{1 - \tilde{\Gamma} \, \phi} \: \tilde{\Gamma} \\
- \frac{1}{2} \: (1+{\bm \sigma} \cdot {\bm
\sigma^{\prime}}) \: \Gamma^\omega \, ({\mathcal G}
{\mathcal G})_{\text{ZS}^\prime} \: \frac{1}{1 -
\Gamma^\omega \, ({\mathcal G} {\mathcal G})_{\text{ZS}^\prime}}
\: \Gamma^\omega .
\label{IIwithIqp}
\end{multline}
The first term $I_{\text{qp}} =
I + \bigl( 1 - \frac{1}{2} \: (1+{\bm \sigma} \cdot {\bm
\sigma^{\prime}}) \bigr) \: \tilde{\Gamma} \, \phi \: (1 -
\tilde{\Gamma} \, \phi)^{-1} \:
\tilde{\Gamma}$ is {\it quasi}particle-{\it quasi}hole
irreducible both in the ZS and ZS$^\prime$ channel. Due to the
identity $P_{\bm \sigma} (1-P_{\bm \sigma}) = - (1-P_{\bm \sigma})$
and the antisymmetry of $I$, $I_{qp}$ is also antisymmetric. For ${\mathbf p}
= {\mathbf p^\prime}$ the non-singular
parts of the ZS and the ZS$^\prime$ graphs differ only in the spin
dependence. This is reflected in the factor $(1-P_{\bm
\sigma})$, which vanishes for a Fermi liquid of say spin up species
only.

Eq. (\ref{IIwithIqp}) is an integral equation for
$\Gamma^\omega$, which diagrammatically is of the
form:
\begin{align}
\hspace{0.9cm}
\parbox[c][4.2cm]{3.6cm}{\begin{fmfgraph*}(20,25)
\fmfpen{1.5}
\fmfleft{p1,pp1}
\fmflabel{$p+\frac{q}{2}$}{p1}
\fmflabel{$p^\prime+\frac{q}{2}$}{pp1}
\fmfright{p2,pp2}
\fmflabel{$p-\frac{q}{2}$}{p2}
\fmflabel{$p^\prime-\frac{q}{2}$}{pp2}
\fmf{fermion}{p1,v4}
\fmf{fermion}{v1,p2}
\fmf{fermion}{v3,pp1}
\fmf{fermion}{pp2,v2}
\fmf{plain}{s1,s2}
\fmfpoly{smooth,tension=0.6}{v1,s1,v2,v3,s2,v4}
\end{fmfgraph*}} \hspace{-0.725cm} &= \hspace{0.875cm}
\parbox[c][4.2cm]{4cm}{\begin{fmfgraph*}(20,25)
\fmfpen{1.5}
\fmfleft{p1,pp1}
\fmflabel{$p+\frac{q}{2}$}{p1}
\fmflabel{$p^\prime+\frac{q}{2}$}{pp1}
\fmfright{p2,pp2}
\fmflabel{$p-\frac{q}{2}$}{p2}
\fmflabel{$p^\prime-\frac{q}{2}$}{pp2}
\fmf{fermion}{p1,v4}
\fmf{fermion}{v1,p2}
\fmf{fermion}{v3,pp1}
\fmf{fermion}{pp2,v2}
\fmf{plain}{s1,s2}
\fmf{plain}{s3,s4}
\fmfpoly{smooth,tension=0.0}{v1,s1,v2,s3,v3,s2,v4,s4}
\end{fmfgraph*}} \hspace{-0.8cm} \nonumber \\
& + \hspace{0.8cm}
\parbox[c][4.5cm]{5.5cm}{\begin{fmfgraph*}(50,30)
\fmfpen{1.5}
\fmfleft{p1,pp1}
\fmflabel{$p+\frac{q}{2}$}{p1}
\fmflabel{$p^\prime+\frac{q}{2}$}{pp1}
\fmfright{p2,pp2}
\fmflabel{$p-\frac{q}{2}$}{p2}
\fmflabel{$p^\prime-\frac{q}{2}$}{pp2}
\fmf{fermion}{p1,v4}
\fmf{fermion}{u1,p2}
\fmf{fermion}{v3,pp1}
\fmf{fermion}{pp2,u2}
\fmf{fermion,right=0.5,tension=0.5,label=$p^{\prime\prime}-\frac{p-p^\prime}{2}$,label.side=right}{u3,v2}
\fmf{fermion,right=0.5,tension=0.5,label=$p^{\prime\prime}+\frac{p-p^\prime}{2}$,label.side=right}{v1,u4}
\fmf{plain}{s1,s2}
\fmf{plain}{s3,s4}
\fmfpoly{smooth,tension=0.4}{v1,v2,s1,v3,v4,s2}
\fmfpoly{smooth,tension=0.4}{u1,u2,s3,u3,u4,s4}
\end{fmfgraph*}} \hspace{0.5cm} \nonumber \\
&+ \hspace{0.6cm}
\parbox[c][4.5cm]{7.5cm}{\begin{fmfgraph*}(80,30)
\fmfpen{1.5}
\fmfleft{p1,pp1}
\fmflabel{$p+\frac{q}{2}$}{p1}
\fmflabel{$p^\prime+\frac{q}{2}$}{pp1}
\fmfright{p2,pp2}
\fmflabel{$p-\frac{q}{2}$}{p2}
\fmflabel{$p^\prime-\frac{q}{2}$}{pp2}
\fmf{fermion}{p1,v4}
\fmf{fermion}{w1,p2}
\fmf{fermion}{v3,pp1}
\fmf{fermion}{pp2,w2}
\fmf{fermion,right=0.5,tension=0.5,label=$p^{\prime\prime}-\frac{p-p^\prime}{2}$,label.side=right}{u3,v2}
\fmf{fermion,right=0.5,tension=0.5,label=$p^{\prime\prime}+\frac{p-p^\prime}{2}$,label.side=right}{v1,u4}
\fmf{fermion,right=0.5,tension=0.5,label=$p^{\prime\prime\prime}-\frac{p-p^\prime}{2}$,label.side=right}{w3,u2}
\fmf{fermion,right=0.5,tension=0.5,label=$p^{\prime\prime\prime}+\frac{p-p^\prime}{2}$,label.side=right}{u1,w4}
\fmf{plain}{s1,s2}
\fmf{plain}{s3,s4}
\fmf{plain}{s5,s6}
\fmfpoly{smooth,tension=0.3}{v1,v2,s1,v3,v4,s2}
\fmfpoly{smooth,tension=0.0}{u1,u2,s3,u3,u4,s4}
\fmfpoly{smooth,tension=0.3}{w1,w2,s5,w3,w4,s6}
\end{fmfgraph*}} + \: \ldots .
\label{II}
\end{align}
The diagram with the crossed lines denotes the
driving term $I_{qp}$, which consists of all quasiparticle-quasihole
irreducible diagrams (in both the ZS and ZS$^\prime$ channels). 
The series of all ZS$^\prime$ bubble diagrams corresponding to the
remaining terms in Eq. (\ref{IIwithIqp}) is called the induced interaction.
It may be regarded as the linear response of the system to the
presence of the quasiparticle. Due to the exchange of the external
lines it is explicit that all diagrams in the induced interaction
are irreducible in the ZS channel. The limit $q \to 0$ is to be
taken only after the iteration of the induced interaction. In order
to illustrate this, consider the one pion exchange vertex function,
Eq. (\ref{OPE}), as driving term. The momentum transfers in
$\Gamma^{\text{OPE}} (p,p^\prime;q)$ are ${\mathbf q}$ and
${\mathbf p} - {\mathbf p^\prime}$. However, due to the exchange
character of the induced interaction, the corresponding momentum
transfers in the vertices of the one ZS$^\prime$ bubble are
${\mathbf p} - {\mathbf p^\prime}$ and $\frac{1}{2}({\mathbf p} +
{\mathbf p^\prime} + {\mathbf q})
- {\mathbf p^{\prime\prime}}$. Although terms proportional
to ${\mathbf q}^2$ in the driving term vanish in the Landau limit,
they appear in the induced interaction. Thus, in
general the induced interaction requires input beyond Fermi liquid
theory, since the Landau parameters are defined only
in the $|{\mathbf q}| \rightarrow 0$ limit. In the limit
${\mathbf p} = {\mathbf p^\prime}$, the induced interaction
expressed solely in terms of Landau parameters is exact. Nevertheless,
applications to nuclear matter~\cite{BBfornuclmat,bigreport,tensor}, neutron
matter~\cite{WAP,Baldo} and liquid $^3$He~\cite{He3.1,He3.2,He3.3} 
have shown that the induced interaction is a very powerful
approximation even for non vanishing angles $\theta$, i.e. ${\mathbf
p} \neq {\mathbf p^\prime}$.

The one ZS$^\prime$ bubble contribution to the induced interactions is given
by
\begin{multline}
\Gamma_{{\bm \sigma} \cdot {\bm \sigma^{\prime}}, {\bm \tau} \cdot
{\bm \tau^{\prime}}}^{\text{ind (2)}} (p,p^\prime) = - \frac{1}{4}
(1+{\bm \sigma} \cdot {\bm \sigma^{\prime}}) (1+{\bm \tau} \cdot
{\bm \tau^{\prime}}) N(0) \: z^2 \\[0.5mm]
\times \frac{1}{4} \: \text{Tr}_{{\bm \sigma^{\prime\prime}}
{\bm \tau^{\prime\prime}}} \int \frac{d \Omega_{{\mathbf
p^{\prime\prime}}}}{4 \pi} \: \Gamma^{\omega}_{{\bm \sigma} \cdot
{\bm \sigma^{\prime\prime}}, {\bm \tau} \cdot {\bm
\tau^{\prime\prime}}}(\frac{p+p^\prime}{2},p^{\prime\prime};p-p^\prime)
\\[1.5mm]
\times \frac{v_{\text{F}} \: \hat{{\mathbf p}}^{\prime\prime} \cdot
({\mathbf p} - {\mathbf p^\prime})}{\varepsilon -
\varepsilon^\prime - v_{\text{F}}
\: \hat{{\mathbf p}}^{\prime\prime} \cdot ({\mathbf p} -
{\mathbf p^\prime})} \: \Gamma^{\omega}_{{\bm \sigma^{\prime\prime}}
\cdot {\bm \sigma^{\prime}}, {\bm \tau^{\prime\prime}} \cdot
{\bm \tau^{\prime}}}(p^{\prime\prime},\frac{p+p^\prime}{2};p-p^\prime) .
\label{II2}
\end{multline}
In the extrapolation away from ${\mathbf p} = {\mathbf p^\prime}$
the initial and final momenta are treated symmetrically. This
yields the correct result, e.g. for a current-current coupling.
Using the bare direct and exchange interaction as driving term,
i.e.
\begin{equation}
\hspace{0.8cm}
\parbox[c][4cm]{3cm}{\begin{fmfgraph*}(20,25)
\fmfpen{1.5}
\fmfleft{p1,pp1}
\fmflabel{$p+\frac{q}{2}$}{p1}
\fmflabel{$p^\prime+\frac{q}{2}$}{pp1}
\fmfright{p2,pp2}
\fmflabel{$p-\frac{q}{2}$}{p2}
\fmflabel{$p^\prime-\frac{q}{2}$}{pp2}
\fmf{fermion}{p1,v4}
\fmf{fermion}{v1,p2}
\fmf{fermion}{v3,pp1}
\fmf{fermion}{pp2,v2}
\fmf{plain}{s1,s2}
\fmf{plain}{s3,s4}
\fmfpoly{smooth,tension=0.0}{v1,s1,v2,s3,v3,s2,v4,s4}
\end{fmfgraph*}} = \hspace{0.5cm}
\parbox[c][4cm]{2.5cm}{\begin{fmfgraph*}(20,25)
\fmfleft{p1,pp1}
\fmfright{p2,pp2}
\fmfforce{(0,0.1h)}{p1}
\fmfforce{(0.3w,0.1h)}{p2}
\fmfforce{(0.7w,0.9h)}{pp1}
\fmfforce{(w,0.9h)}{pp2}
\fmfforce{(0.15w,0.5h)}{v1}
\fmfforce{(0.85w,0.5h)}{v2}
\fmf{fermion}{p1,v1,p2}
\fmf{dashes}{v1,v2}
\fmf{fermion}{pp2,v2,pp1}
\end{fmfgraph*}} + \hspace{0.5cm}
\parbox[c][4cm]{2.5cm}{\begin{fmfgraph*}(20,25)
\fmfleft{p1,pp1}
\fmfright{p2,pp2}
\fmfforce{(0,0.1h)}{p1}
\fmfforce{(w,0.1h)}{p2}
\fmfforce{(0,0.9h)}{pp1}
\fmfforce{(w,0.9h)}{pp2}
\fmfforce{(0.15w,0.5h)}{v1}
\fmfforce{(0.85w,0.5h)}{v2}
\fmf{fermion}{p1,v1,pp1}
\fmf{dashes}{v1,v2}
\fmf{fermion}{pp2,v2,p2}
\end{fmfgraph*}} ,
\label{bareint}
\end{equation}
one finds that the lowest order contributions to Eq. (\ref{II2})
correspond to the following diagrams
\begin{equation}
\Gamma^{\text{{ind} (2)}} = \hspace{0.2cm}
\parbox[c][2.8cm]{2.7cm}{\begin{fmfgraph*}(25,25)
\fmfleft{p1,pp1}
\fmfright{p2,pp2}
\fmfforce{(0,0)}{p1}
\fmfforce{(w,0)}{p2}
\fmfforce{(0,h)}{pp1}
\fmfforce{(w,h)}{pp2}
\fmfforce{(0.1w,0.3h)}{v1}
\fmfforce{(0.9w,0.7h)}{v2}
\fmfforce{(0.5w,0.3h)}{v3}
\fmfforce{(0.5w,0.7h)}{v4}
\fmf{fermion}{p1,v1,pp1}
\fmf{dashes}{v1,v3}
\fmf{dashes}{v2,v4}
\fmf{fermion}{pp2,v2,p2}
\fmf{fermion,left=0.5,tension=0.3}{v3,v4}
\fmf{fermion,left=0.5,tension=0.3}{v4,v3}
\end{fmfgraph*}} \hspace{0.1cm} + \hspace{0.1cm}
\parbox[c][2.8cm]{2.4cm}{\begin{fmfgraph*}(22.5,25)
\fmfleft{p1,pp1}
\fmfright{p2,pp2}
\fmfforce{(0,0)}{p1}
\fmfforce{(w,0)}{p2}
\fmfforce{(0,h)}{pp1}
\fmfforce{(w,h)}{pp2}
\fmfforce{(0.2w,0.3h)}{v1}
\fmfforce{(0.8w,0.3h)}{v2}
\fmfforce{(0.2w,0.7h)}{v3}
\fmfforce{(0.8w,0.7h)}{v4}
\fmf{dashes}{v1,v3}
\fmf{dashes}{v2,v4}
\fmf{fermion}{p1,v1,v2,p2}
\fmf{fermion}{pp2,v4,v3,pp1}
\end{fmfgraph*}} + \hspace{0.1cm}
\parbox[c][2.8cm]{2.5cm}{\begin{fmfgraph*}(22.5,25)
\fmfleft{p1,pp1}
\fmfright{p2,pp2}
\fmfforce{(0,0)}{p1}
\fmfforce{(w,0)}{p2}
\fmfforce{(0,h)}{pp1}
\fmfforce{(w,h)}{pp2}
\fmfforce{(0.5w,0.5h)}{v1}
\fmfforce{(0.9w,0.5h)}{v2}
\fmfforce{(0.25w,0.25h)}{v3}
\fmfforce{(0.25w,0.75h)}{v4}
\fmf{dashes}{v1,v2}
\fmf{fermion}{pp2,v2,p2}
\fmf{fermion}{p1,v3,v1,v4,pp1}
\fmf{dashes}{v3,v4}
\end{fmfgraph*}} + \hspace{0.2cm}
\parbox[c][2.8cm]{2.5cm}{\begin{fmfgraph*}(22.5,25)
\fmfforce{(0,0)}{p1}
\fmfforce{(w,0)}{p2}
\fmfforce{(0,h)}{pp1}
\fmfforce{(w,h)}{pp2}
\fmfforce{(0.5w,0.5h)}{v1}
\fmfforce{(0.1w,0.5h)}{v2}
\fmfforce{(0.75w,0.25h)}{v3}
\fmfforce{(0.75w,0.75h)}{v4}
\fmf{dashes}{v1,v2}
\fmf{fermion}{p1,v2,pp1}
\fmf{fermion}{pp2,v4,v1,v3,p2}
\fmf{dashes}{v3,v4}
\end{fmfgraph*}} \hspace{-0.2cm}.
\label{diag2ndorder}
\end{equation}

We expand the angular dependence of the quasiparticle interaction
$\Gamma^\omega$ on Legendre polynomials, $\Gamma^\omega = \sum_l
\Gamma^\omega_l \, P_l \, (\cos \theta)$. After inserting this in
Eq. (\ref{II2}), we find
\begin{multline}
\Gamma_{{\bm \sigma} \cdot {\bm \sigma^{\prime}}, {\bm \tau} \cdot
{\bm \tau^{\prime}}}^{\text{ind (2)}} (p,p^\prime) = - \frac{1}{4}
(1+{\bm \sigma} \cdot {\bm \sigma^{\prime}}) (1+{\bm \tau} \cdot
{\bm \tau^{\prime}}) N(0) \: z^2 \\[1mm]
\times \frac{1}{4} \: \text{Tr}_{{\bm \sigma^{\prime\prime}} {\bm
\tau^{\prime\prime}}} \sum_{l, l^\prime} \Gamma^{\omega}_{{\bm
\sigma} \cdot {\bm \sigma^{\prime\prime}}, {\bm \tau} \cdot {\bm
\tau^{\prime\prime}}, \: l} \: \Gamma^{\omega}_{{\bm
\sigma^{\prime\prime}} \cdot {\bm \sigma^{\prime}},
{\bm \tau^{\prime\prime}} \cdot {\bm \tau^{\prime}}, \: l^\prime} \\
\times \int \frac{d \Omega_{{\mathbf p^{\prime\prime}}}}{4 \pi} \:
P_l(\frac{\widehat{{\mathbf p} + {\mathbf p^\prime}}}{2} \cdot
\hat{{\mathbf p}}^{\prime\prime}) \; P_{l^\prime} (\hat{{\mathbf
p}}^{\prime\prime} \cdot \frac{\widehat{{\mathbf p} + {\mathbf
p}^\prime}}{2}) \; \frac{v_{\text{F}} \: \hat{{\mathbf
p}}^{\prime\prime} \cdot ({\mathbf p} - {\mathbf
p^\prime})}{\varepsilon - \varepsilon^\prime - v_{\text{F}} \:
\hat{{\mathbf p}}^{\prime\prime} \cdot ({\mathbf p} -
{\mathbf p^\prime})} \, .
\label{II3}
\end{multline}
In order to cover all possible combinations of ${\mathbf p}$ and
${\mathbf p^\prime}$, the induced interaction is needed for
momentum transfer ${\mathbf q^\prime} = {\mathbf p} - {\mathbf
p^\prime}$ up to $2 k_{\text{F}}$. This is done by extrapolating
the quasiparticle-quasihole propagator in Eq. (\ref{PP}) to large
$q$ using the particle-hole propagator of a free Fermi gas with an
effective mass $m^\star$. Furthermore, the external quasiparticles
are assumed to be on the Fermi surface, so that $\varepsilon =
\varepsilon^\prime = 0$. For $l,l^\prime = 0,1$ the resulting integrals in
Eq. (\ref{II3}) are given in~\cite{BBfornuclmat}.
We introduce the notation ${\mathcal F}_{\text{ind}} = N(0) \, z^2 \,
\Gamma_{\text{ind}}$ and decompose the induced interaction into its
scalar, spin, isospin and spin-isospin components,
\begin{equation}
{\mathcal F}_{\text{ind}} = F_{\text{ind}} + F^{\prime}_{\text{ind}} \: {\bm \tau} \cdot {\bm \tau^{\prime}} + G_{\text{ind}} \: {\bm \sigma} \cdot {\bm \sigma^{\prime}} + G^{\prime}_{\text{ind}} \: {\bm \tau} \cdot {\bm \tau^{\prime}} \: {\bm \sigma} \cdot {\bm \sigma^{\prime}} .
\end{equation}
The resulting expression for the scalar induced
interaction, $F_{\text{ind}}$, including $l=0,1$ terms,
is~\cite{BBfornuclmat,bigreport}
\begin{align}
4 F_{\text{ind}} &= 1 \cdot \biggl( \frac{F_0^2 \alpha_0(q^\prime/k_{\text{F}})}{1+F_0 \alpha_0(q^\prime/k_{\text{F}})} + (1-\frac{q^{\prime 2}}{4 k_{\text{F}}^2}) \frac{F_1^2 \alpha_1(q^\prime/k_{\text{F}})}{1+F_1 \alpha_1(q^\prime/k_{\text{F}})} \biggr) \nonumber \\
&+ 3 \cdot \biggl( \frac{F_0^{\prime 2} \alpha_0(q^\prime/k_{\text{F}})}{1+F_0^{\prime} \alpha_0(q^\prime/k_{\text{F}})} + (1-\frac{q^{\prime 2}}{4 k_{\text{F}}^2}) \frac{F_1^{\prime 2} \alpha_1(q^\prime/k_{\text{F}})}{1+F_1^{\prime} \alpha_1(q^\prime/k_{\text{F}})} \biggr) \nonumber \\
&+ 3 \cdot \biggl( \frac{G_0^2 \alpha_0(q^\prime/k_{\text{F}})}{1+G_0 \alpha_0(q^\prime/k_{\text{F}})} + (1-\frac{q^{\prime 2}}{4 k_{\text{F}}^2}) \frac{G_1^2\alpha_1(q^\prime/k_{\text{F}})}{1+G_1 \alpha_1(q^\prime/k_{\text{F}})} \biggr) \nonumber \\
&+ 9 \cdot \biggl( \frac{G_0^{\prime 2}
\alpha_0(q^\prime/k_{\text{F}})}{1+G_0^{\prime}
\alpha_0(q^\prime/k_{\text{F}})} + (1-\frac{q^{\prime 2}}{4
k_{\text{F}}^2}) \frac{G_1^{\prime 2}
\alpha_1(q^\prime/k_{\text{F}})}{1+G_1^{\prime}
 \alpha_1(q^\prime/k_{\text{F}})} \biggr) \, ,
\label{Find}
\end{align}
where $q^\prime = |{\mathbf q^\prime}|$, $\alpha_0(x)$ and
$\alpha_1(x)$ are the Lindhard (or density-density) and
current-current correlation functions, respectively. The
factor $(1-q^{\prime 2}/4 k_{\text{F}}^2)$
guarantees that the current response vanishes for back to back
scattering.
\begin{align}
\alpha_0(x) &= \frac{1}{2} + \frac{1}{2} \biggl( \frac{x}{4} -
\frac{1}{x} \biggr) \ln \frac{1-x/2}{1+x/2} \\
\alpha_1(x) &= \frac{1}{2} \biggl[\frac{3}{8} - \frac{1}{2x^2}+ \biggl(\frac{1}{2x^3}+\frac{1}{4x}-\frac{3x}{32} \biggr) \ln \frac{1+x/2}{1-x/2} \biggr]
\end{align}
For the spin, isospin and spin-isospin induced parts, the coefficients
in (\ref{Find}) have to be changed according to the table below. These
coefficients follow from the recoupling of
spin and isospin between the two particle-hole channels.
\begin{center}
\begin{tabular}{@{\hspace{0.5cm}}c@{\hspace{0.5cm}}|@{\hspace{0.5cm}}c@{\hspace{0.5cm}}|@{\hspace{0.5cm}}c@{\hspace{0.5cm}}|@{\hspace{0.5cm}}c@{\hspace{0.5cm}}|@{\hspace{0.5cm}}c@{\hspace{0.5cm}}}
& $F$ & $F^\prime$ & $G$ & $G^\prime$ \\
\hline
$F_{\text{ind}}$ & 1 & 3 & 3 & 9 \\
\hline
$F_{\text{ind}}^\prime$ & 1 & -1 & 3 & -3 \\
\hline
$G_{\text{ind}}$ & 1 & 3 & -1 & -3 \\
\hline
$G_{\text{ind}}^\prime$ & 1 & -1 & -1 & 1 \\
\end{tabular}
\end{center}
By construction, the induced interaction with the bare direct and
exchange interaction as driving term generates the complete
particle-hole parquet for the scattering amplitude. The
particle-hole parquet are all planar fermionic diagrams except
those that are joined by the particle-particle (BCS) channel. This
corresponds to the solution to the fermionic parquet equations of
Lande and Smith ignoring the coupling to the \textbf{s}
channel~\cite{parquet}. The ${\mathbf s}$ channel diagrams are
particle-hole irreducible in both the ZS and ZS$^\prime$ channels.
Hence, they should be included in the driving term. Traditionally
the driving term has been computed within Brueckner theory by
varying the energy twice with respect to the occupation number and
removing all contributions that are included in the induced
interaction~\cite{BBfornuclmat,dterm}. If two hole contributions,
which are expected to be small, are neglected, one can express the
driving term as the direct and exchange Brueckner $G$ matrix
multiplied by the renormalization factor $z^2$. The factor $z^2$
accounts for some of the higher order
completely particle-hole irreducible diagrams. Diagrams involving
e.g. particle-particle ladders with a screened interaction are
neglected. In this work we identify the $G$ matrix in the driving
term with the low momentum nucleon-nucleon interaction
$V_{\text{low k}}$ for $\Lambda = k_{\text{F}}$~\cite{Vlowkflow}
and consequently employ $z^2 \, V_{\text{low k}}$ as the driving
term. As discussed in the introduction, $V_{\text{low
k}}$ is smooth and includes the effects of the short range
repulsion.

\section{Relation between the Fermi Liquid Parameters and the Low Momentum Nucleon-Nucleon Interaction}

For ${\mathbf p} = {\mathbf p^\prime}$ the induced interaction
expressed in terms of Fermi liquid parameters is exact and one can
derive general constraints for these parameters. In
this limit the integral in Eq. (\ref{II3}) simplifies to $-
\delta_{l,l^\prime} / (2l+1)$ and all higher ZS$^\prime$ bubble
terms are easily summed. Thus, Eq. (\ref{II}) can be written as
follows
\begin{multline}
F_s + F_a \: {\bm \sigma} \cdot {\bm \sigma^{\prime}} = F_s^d + F_a^d \:
{\bm \sigma} \cdot {\bm \sigma^{\prime}} + \int \frac{d
\Omega_{{\mathbf p^{\prime\prime}}}}{4 \pi} \: \bigl\{ F_s
(p,p^{\prime\prime}) A_s (p^{\prime\prime},p) \: \frac{1 + {\bm \sigma} \cdot
{\bm \sigma^{\prime}}}{2} \\
+ F_a (p,p^{\prime\prime}) A_a
(p^{\prime\prime},p) \: \frac{3 - {\bm \sigma} \cdot {\bm
\sigma^{\prime}}}{2} \bigr\} \, ,
\label{inteq1}
\end{multline}
where we again consider a system with spin only. For
${\mathbf p} = {\mathbf p^\prime}$ the series of ZS$^\prime$ bubbles
is equivalent to the series of ZS bubbles summed by the scattering
amplitude up to a sign and the spin exchange operator for the exchange
of the external lines in the induced interaction. The
equation for the scattering amplitude reads
\begin{multline}
A_s + A_a \: {\bm \sigma} \cdot {\bm \sigma^{\prime}} = F_s + F_a \:
{\bm \sigma} \cdot {\bm \sigma^{\prime}} - \int \frac{d
\Omega_{{\mathbf p^{\prime\prime}}}}{4 \pi} \: \bigl\{ F_s
(p,p^{\prime\prime}) A_s (p^{\prime\prime},p) \\
+ F_a (p,p^{\prime\prime}) A_a
(p^{\prime\prime},p) \: {\bm \sigma} \cdot {\bm \sigma^{\prime}}
\bigr\} \, .
\label{inteq2}
\end{multline}
We have introduced the notation
\begin{align}
\mathcal{F} &= F_s + F_a \: {\bm \sigma} \cdot {\bm \sigma^{\prime}}
\\
\mathcal{A} &= A_s + A_a \: {\bm \sigma} \cdot {\bm \sigma^{\prime}}
\\
\mathcal{F_{\text{driving}}} &= F_s^d + F_a^d \: {\bm \sigma} \cdot
{\bm \sigma^{\prime}} .
\end{align}
It is easy to solve the integral equations for the
driving term. In the $S=1$ channel the sum and in the $S=0$ channel
the difference of the two integral equations, Eqs. (\ref{inteq1})
and (\ref{inteq2}), leads to
\begin{align}
\mathcal{F_{\text{driving}}} \: (S=1) \: &= \: \mathcal{A} \: = \: 0
\label{constS0} \\
\mathcal{F_{\text{driving}}} \: (S=0) \: &= \: 2 {\mathcal F} -
{\mathcal A} \, ,
\label{constS1}
\end{align}
where we have used the Pauli principle sum rule in the case $S=1$.
The first case, Eq. (\ref{constS0}), projects on odd partial waves,
while the second case, Eq. (\ref{constS1}), projects on even
partial waves. In symmetric nuclear matter, there are two
spin-isospin states corresponding to odd partial waves ($S=T=0$ and
$S=T=1$) and two corresponding to even states ($S=0, \, T=1$ and $S=1, \,
T=0$). We thus obtain two new constraints on the
Fermi liquid parameters of nuclear matter:
\begin{align}
& \sum_{l} \biggl\{ 2 F_l - \frac{F_l}{1+F_l/(2l+1)} \: + \: 2
F_l^{\prime}  - \frac{F_l^{\prime}}{1+F_l^{\prime}/(2l+1)} \nonumber \\
& \hspace{2cm} -3 \: \bigl( 2 G_l - \frac{G_l}{1+G_l/(2l+1)} \bigr) \:
- 3 \: \bigl( 2 G_l^{\prime} -
\frac{G_l^{\prime}}{1+G_l^{\prime}/(2l+1)} \bigr) \biggr\} \nonumber
\\[1mm]
& \hspace{4.2cm} = \: \mathcal{F_{\text{driving}}} \: (S=0, T=1) \label{C1} \\
& \sum_{l} \biggl\{ 2 F_l - \frac{F_l}{1+F_l/(2l+1)} \: -3 \: \bigl( 2
F_l^{\prime}  - \frac{F_l^{\prime}}{1+F_l^{\prime}/(2l+1)} \bigr) \nonumber \\
& \hspace{2cm} + \: 2 G_l - \frac{G_l}{1+G_l/(2l+1)} \: -3 \: \bigl( 2
G_l^{\prime} - \frac{G_l^{\prime}}{1+G_l^{\prime}/(2l+1)} \bigr)
\biggr\} \nonumber \\[1mm]
& \hspace{4.2cm} = \: \mathcal{F_{\text{driving}}} \: (S=1, T=0) .
\label{C2}
\end{align}
These are general constraints, which however are useful only if the
driving term is known. Such a constraint was first derived by Bedell
and Ainsworth~\cite{He3.2} for paramagnetic Fermi liquids, like liquid
$^3$He or $^3$He$\,-^4$He mixtures, and employed to extract the
effective scattering length. As reasoned above, we
approximate the driving term with $z^2 \, V_{\text{low k}}$. We need
the matrix elements of $V_{\text{low k}}$ in the basis of total
spin $S$ and total isospin $T$. By summing over $M_S$, we project
onto the central components of the forward scattering
amplitude~\cite{tensor}.
\begin{equation}
\frac{1}{V} \: \mathcal{F_{\text{driving}}} \: (S,T)
= \frac{z^2}{2 S + 1} \: N(0) \, \sum_{M_S}
\biggl( \langle {\mathbf p} \: {\mathbf p^\prime} \: S \: T |
V_{\text{low k}} | {\mathbf p} \: {\mathbf p^\prime} \: S \: T \rangle -
\text{exchange} \biggr) \, .
\end{equation}
Transforming to relative momentum ${\mathbf q^\prime} = {\mathbf p} -
{\mathbf p^\prime}$ and coupling angular momentum and total spin leads to
\begin{multline}
\mathcal{F_{\text{driving}}} \: (S,T) =
z^2 \, N(0) \: \frac{4 \pi}{2 S + 1} \: \sum_{J \: ,
\: l} \: ( 2 J + 1 ) \: \bigl( 1 - (-1)^{l+S+T} \bigr) \\
\times \langle k = \frac{q^\prime}{2}
\: l \: S \: J \: T |
V_{\text{low k}} | k = \frac{q^\prime}{2} \: l \: S \: J
\: T \rangle \, .
\end{multline}
At vanishing relative momentum there are only s-wave
contributions to the driving term due to the rotational invariance.
Since the driving term is antisymmetric, these contributions are in
the $S=0$, $T=1$ and $S=1$, $T=0$ channels, consistent with the two
Pauli principle sum rules. Thus, with the input $z^2 \, V_{\text{low
k}}$ for the driving term, the two relations, Eqs. (\ref{C1}) and
(\ref{C2}), constrain the dimensionless Fermi liquid parameters of
nuclear matter in a nontrivial way to
\begin{align}
\mathcal{F_{\text{driving}}} \: (S=0, T=1) & = z^2 \, \frac{16 \,
m_{\text{N}} \, k_{\text{F}} (1+F_1/3)}{\pi} \, V_{\text{low k}}
(0,0;\Lambda=k_{\text{F}},{}^1\text{S}_0) \\
\mathcal{F_{\text{driving}}} \: (S=1, T=0) & =
z^2 \, \frac{16 \, m_{\text{N}} \, k_{\text{F}}
(1+F_1/3)}{\pi} \, V_{\text{low k}}
(0,0;\Lambda=k_{\text{F}},{}^3\text{S}_1) .
\end{align}
The dimension of the
potential is absorbed by the density of states. As explained in the
introduction and in~\cite{Vlowkflow}, $V_{\text{low k}}$ is
obtained from a RG decimation of various nuclear force models.
Bogner {\it et al.}~\cite{Vlowkflow} find that the $V_{\text{low
k}}$ obtained from various bare potentials at $\Lambda =
k_{\text{F}}$ are identical. Moreover, when one compares the low
momentum part of the bare interaction models with $V_{\text{low
k}}$, one observes that the main effect of the renormalization
is a constant shift in momentum space. This correspond to a
smeared delta function in coordinate space and accounts for the
removal of the model dependent short range core. Thus, the two constraints,
which use as dynamical input $V_{\text{low k}} (0,0)$, connect the
pivotal matrix element of the RG decimation to the unique set of
Fermi liquid parameters of nuclear matter. As the Fermi liquid
parameters are fixed points under the RG flow towards the Fermi
surface, the constraints relate $V_{\text{low k}}$ to these
fixed points.

\section{Renormalization Group with the Induced Interaction}

In the microscopic derivation of Fermi liquid theory, one isolates
the quasiparticle part of the full propagator from the pair
background. We have shown that, for ${\mathbf p}\approx {\mathbf
p^\prime}$, this is rigorously possible also when
both particle-hole channels are taken into account. This is
necessary in order to preserve the Pauli principle and leads to the
induced interaction. Having reduced the theory to
interactions among quasiparticles, we now separate the soft modes
of the quasiparticle-quasihole propagators from the hard ones. To
this end we introduce a momentum cutoff at
$k_{\text{F}}\pm\Lambda_{\text{F}}$~\footnote{We denote with
$\Lambda_{\text{F}}$ the cutoff in medium, which is not to be
confused with the cutoff $\Lambda$ for $V_{\text{low k}}$.}. In
this way we arrive at a theory of quasiparticles interacting in
a model space of slow modes exclusively. For a
discussion of the RG approach to Fermi liquid theory
see~\cite{FLTandRG1,FLTandRG2,FLTandRG3}.

In a shorthand notation we write $({\mathcal G}
{\mathcal G})_{\text{ZS}} = ({\mathcal G} {\mathcal
G})_{\text{ZS}}^{\text{S}} + ({\mathcal G} {\mathcal
G})_{\text{ZS}}^{\text{H}}$ for the ZS propagators (at finite $q$)
and with analogous expressions for the ZS$^\prime$ channel. The
indices S and H denote integrations over the soft
(inside the shell) and hard (outside) momenta, respectively. We
define the vertices $\gamma^q (p,p^\prime;q,\Lambda_{\text{F}})$
and $\gamma^\omega (p,p^\prime;q,\Lambda_{\text{F}})$ by
\begin{align}
\gamma^q_{\text{ZS}} (\Lambda_{\text{F}}) & = \gamma^\omega_{\text{ZS}}
(\Lambda_{\text{F}}) + \gamma^\omega_{\text{ZS}}
(\Lambda_{\text{F}}) \: ({\mathcal G} {\mathcal
G})_{\text{ZS}}^{\text{H}} \: \gamma^q_{\text{ZS}}
(\Lambda_{\text{F}}) \label{SAwithLa} \\[1.5mm]
\gamma^\omega_{\text{ZS}} 
(\Lambda_{\text{F}}) & = I_{\text{qp}} - \: \bigl\{
\gamma^\omega_{\text{ZS}^\prime} (\Lambda_{\text{F}}) \:
({\mathcal G} {\mathcal G})_{\text{ZS}^\prime}^{\text{H}} \:
\gamma^\omega_{\text{ZS}^\prime} (\Lambda_{\text{F}}) \nonumber \\[1mm]
& \hspace{1cm} + \gamma^\omega_{\text{ZS}^\prime} (\Lambda_{\text{F}}) \:
({\mathcal G} {\mathcal G})_{\text{ZS}^\prime}^{\text{H}} \:
\gamma^\omega_{\text{ZS}^\prime} (\Lambda_{\text{F}}) \: 
({\mathcal G} {\mathcal G})_{\text{ZS}^\prime}^{\text{H}} \:
\gamma^\omega_{\text{ZS}^\prime} 
(\Lambda_{\text{F}}) + \ldots \bigr\} \label{IIwithLa} \\[1.5mm]
& = I_{\text{qp}} - \gamma^\omega_{\text{ZS}^\prime}
(\Lambda_{\text{F}}) \: ({\mathcal G} {\mathcal
G})_{\text{ZS}^\prime}^{\text{H}} \: \gamma^q_{\text{ZS}^\prime}
(\Lambda_{\text{F}}) \, , 
\label{II2withLa}
\end{align}
where $I_{\text{qp}}$ denotes the quasiparticle-quasihole
irreducible driving term defined above. Furthermore, we introduce the
shorthand notation $\gamma^q_{\text{ZS}} = \gamma^q
(p,p^\prime;q,\Lambda_{\text{F}})$ and the exchange thereof 
$\gamma^q_{\text{ZS}^\prime} = \gamma^q
((p+p^\prime+q)/2,(p+p^\prime-q)/2;p-p^\prime,\Lambda_{\text{F}})$,
where for simplicity the spin- and isospin-dependence is
suppressed. Analogous expressions hold for $\gamma^\omega_{\text{ZS}}$
and $\gamma^\omega_{\text{ZS}^\prime}$. Due
to phase space restrictions, the running of $\gamma^q$ and
$\gamma^\omega$ at $T=0$ starts at $\Lambda_{\text{F}} = \max ( |{\mathbf
q}|/2 , |{\mathbf p} - {\mathbf p^\prime}|/2 )$. With a weakly energy 
dependent driving term $I_{qp}$, we can set $\omega = 0$ in the flow
equations. For $\Lambda_{\text{F}} = 0$, the quasiparticle scattering 
amplitude $\Gamma^q$ and the quasiparticle interaction $\Gamma^\omega$
are obtained as the $|{\mathbf q}| \rightarrow 0$ limit of 
$\gamma^q_{\text{ZS}}$ and $\gamma^\omega_{\text{ZS}}$, 
respectively. On the other hand, for $\Lambda_{\text{F}} 
\geq k_{\text{F}}$, the particle-hole contributions vanish in 
the momentum range of interest
$|{\mathbf q}|,|{\mathbf p} - {\mathbf p^\prime}| \leq 2
k_{\text{F}}$, so that
$\gamma^q_{ZS}(k_{\text{F}}) = \gamma^\omega_{ZS}(k_{\text{F}}) = I_{qp}$.

We differentiate Eqs. (\ref{SAwithLa}) and
(\ref{IIwithLa}) with respect to $\Lambda_{\text{F}}$ and require
$d I_{\text{qp}}/d
\Lambda_{\text{F}}=0$. This corresponds to ignoring the flow from
the particle-particle (BCS) channel. The coupled RG equations then read:
\begin{align}
\frac{d \gamma^q_{\text{ZS}}}{d \Lambda_{\text{F}}} & =
\gamma^q_{\text{ZS}} \: \frac{d ({\mathcal G} 
{\mathcal G})_{\text{ZS}}^{\text{H}}}{d \Lambda_{\text{F}}}
\: \gamma^q_{\text{ZS}} + \frac{d \gamma^\omega_{\text{ZS}}}{d 
\Lambda_{\text{F}}} +
\frac{d \gamma^\omega_{\text{ZS}}}{d \Lambda_{\text{F}}} \:
({\mathcal G} {\mathcal G})_{\text{ZS}}^{\text{H}} \: \gamma^q_{\text{ZS}}
\nonumber \\[1mm]
& \hspace{2.6cm} + \gamma^q_{\text{ZS}} \:
({\mathcal G} {\mathcal G})_{\text{ZS}}^{\text{H}} \: \frac{d
\gamma^\omega_{\text{ZS}}}{d \Lambda_{\text{F}}} + \gamma^q_{\text{ZS}} \:
({\mathcal G} {\mathcal G})_{\text{ZS}}^{\text{H}} \: \frac{d
\gamma^\omega_{\text{ZS}}}{d \Lambda_{\text{F}}} \: ({\mathcal G} {\mathcal
G})_{\text{ZS}}^{\text{H}} \: \gamma^q_{\text{ZS}} \label{SAwithLa2} \\[1.5mm]
\frac{d \gamma^\omega_{\text{ZS}}}{d \Lambda_{\text{F}}} & = - \:
\biggl\{ \frac{1}{1 - \gamma^\omega_{\text{ZS}^\prime} \: 
({\mathcal G} {\mathcal
G})_{\text{ZS}^\prime}^{\text{H}}}
\: \frac{d \gamma^\omega_{\text{ZS}^\prime}}{d \Lambda_{\text{F}}} + 
\frac{1}{1 - \gamma^\omega_{\text{ZS}^\prime} \: ({\mathcal G} 
{\mathcal G})_{\text{ZS}^\prime}^{\text{H}}} \:
\bigl( \gamma^\omega_{\text{ZS}^\prime} \: \frac{d
({\mathcal G} {\mathcal G})_{\text{ZS}^\prime}^{\text{H}}}{d
\Lambda_{\text{F}}} \nonumber \\[1mm]
& \hspace{2.6cm} + \frac{d \gamma^\omega_{\text{ZS}^\prime}}{d
\Lambda_{\text{F}}} \: ({\mathcal G} 
{\mathcal G})_{\text{ZS}^\prime}^{\text{H}} \bigr) \frac{1}{1 -
\gamma^\omega_{\text{ZS}^\prime} \: ({\mathcal G} 
{\mathcal G})_{\text{ZS}^\prime}^{\text{H}}} \:
\gamma^\omega_{\text{ZS}^\prime} - \frac{d
\gamma^\omega_{\text{ZS}^\prime}}{d \Lambda_{\text{F}}} \biggr\} \, .
\label{IIwithLa2}
\end{align}
Using the notation 
\begin{multline}
\delta_{\text{ZS}}
(\Lambda_{\text{F}}) =  \frac{d \gamma^\omega_{\text{ZS}}}{d
\Lambda_{\text{F}}} \: ({\mathcal G} 
{\mathcal G})_{\text{ZS}}^{\text{H}} \: \gamma^q_{\text{ZS}}
+ \gamma^q_{\text{ZS}} \: ({\mathcal G} 
{\mathcal G})_{\text{ZS}}^{\text{H}} \: 
\frac{d \gamma^\omega_{\text{ZS}}}{d
\Lambda_{\text{F}}} \\ + \gamma^q_{\text{ZS}} \:
({\mathcal G} {\mathcal G})_{\text{ZS}}^{\text{H}} \: \frac{d
\gamma^\omega_{\text{ZS}}}{d \Lambda_{\text{F}}} \: ({\mathcal G} {\mathcal
G})_{\text{ZS}}^{\text{H}} \: \gamma^q_{\text{ZS}} \, ,
\end{multline}
and the analogous expression for the ZS$^\prime$ channel, we write
the RG equations in the compact form
\begin{align}
\frac{d \gamma^q_{\text{ZS}}}{d \Lambda_{\text{F}}} & =  
\gamma^q_{\text{ZS}} \: \frac{d
({\mathcal G} {\mathcal G})_{\text{ZS}}^{\text{H}}}{d \Lambda_{\text{F}}}
\: \gamma^q_{\text{ZS}} + \frac{d \gamma^\omega_{\text{ZS}}}{d
\Lambda_{\text{F}}} + \delta_{\text{ZS}} 
(\Lambda_{\text{F}}) \label{RG1} \\[1.5mm]
\frac{d \gamma^\omega_{\text{ZS}}}{d \Lambda_{\text{F}}} & = - \:
\bigl\{ \gamma^q_{\text{ZS}^\prime} \: \frac{d ({\mathcal G} {\mathcal
G})_{\text{ZS}^\prime}^{\text{H}}}{d \Lambda_{\text{F}}} \:
\gamma^q_{\text{ZS}^\prime} + \delta_{\text{ZS}^\prime}
(\Lambda_{\text{F}}) \bigr\} \, .
\label{RG2}
\end{align}
In the limit ${\mathbf p} = {\mathbf p^\prime}$ we can
replace $\gamma^q_{\text{ZS}^\prime}$ in Eq. (\ref{RG2}) by
$\gamma^q_{\text{ZS}}$ and obtain
\begin{equation}
\frac{d \gamma^\omega_{\text{ZS}}}{d \Lambda_{\text{F}}} = - P_{\bm
\sigma} \, \bigl\{ \gamma^q_{\text{ZS}} \: \frac{d ({\mathcal
G} {\mathcal
G})_{\text{ZS}}^{\text{H}}}{d \Lambda_{\text{F}}} \:
\gamma^q_{\text{ZS}} + \delta_{\text{ZS}}
(\Lambda_{\text{F}}) \bigr\} \, ,
\end{equation} 
where the spin structure in the exchange channel is accounted for
by $P_{\bm \sigma}$. This implies that for
${\mathbf p} = {\mathbf p^\prime}$ we have $d \gamma^q_{\text{ZS}} 
| _{|{\mathbf q}|=0} / d \Lambda_{\text{F}} = 0$ in singlet-odd 
and triplet-odd states, while $d ( 2 \gamma^\omega_{\text{ZS}}  - 
\gamma^q_{\text{ZS}} ) | _{|{\mathbf q}|=0} / d
\Lambda_{\text{F}} = 0$ in singlet-even and triplet-even states.
Thus, the Pauli principle sum rules and the new constraints are
invariant under the RG flow. The coupled RG equations, 
Eqs. (\ref{SAwithLa2}) and
(\ref{IIwithLa2}), are nonperturbative. To lowest order, where
$\delta$ in Eqs. (\ref{RG1}) and (\ref{RG2}) is neglected, these
agree with the perturbative one-loop RG equations of
Dupuis~\cite{FLTandRG3}.

\section{Fermi Liquid Parameters and Tensor Interactions}

The aim of this section is to study whether phenomenological values
for the Fermi liquid parameters are consistent with the sum rules
as well as the constraints. For this purpose we approximate the
quasiparticle interaction by the $l=0$ and $l=1$ terms. 
As additional input we take the
phenomenological values for the scalar and isospin Fermi liquid
parameters. The central spin and spin-isospin Fermi liquid
parameters are then obtained from the sum rules and
the constraints. By taking linear combinations of the sum rules
Eqs. (\ref{SR1}) and (\ref{SR2}), and the constraints, Eqs.
(\ref{C1}) and (\ref{C2}), the equations for $G_l$ and $G_l^\prime$
decouple:
\begin{align}
&\sum_{l} \biggl\{ \frac{F_l}{1 + F_l/(2 l + 1)} + 3 \:
\frac{G_l^{\prime}}{1 + G_l^{\prime} / (2 l + 1)} \biggr\} = 0
\label{eq1} \\
&\sum_{l} \biggl\{ 2 F_l - \frac{F_l}{1+F_l/(2l+1)} \: - 2 \: \bigl( 2
F_l^{\prime}  - \frac{F_l^{\prime}}{1+F_l^{\prime}/(2l+1)} \bigr)
\nonumber \\
& \hspace{0.5cm} - 3 \: \bigl( 2 G_l^{\prime} -
\frac{G_l^{\prime}}{1+G_l^{\prime}/(2l+1)} \bigr)
\biggr\} \nonumber \\
& \hspace{1cm} = \: z^2 \: \frac{16 \, m_{\text{N}}  \, k_{\text{F}}
(1+F_1/3)}{\pi} \; V_{\text{low k}}
(0,0;\Lambda=k_{\text{F}},\frac{^1\text{S}_0+3\cdot {}^3\text{S}_1}{4})
\label{eq2} \\
&\sum_{l} \biggl\{ \frac{F_l}{1 + F_l / (2 l + 1)} + \frac{3}{2} \:
\frac{F_l^{\prime}}{1 + F_l^{\prime} / (2 l + 1)} + \frac{3}{2} \:
\frac{G_l}{1 + G_l / (2 l + 1)} \biggr\} = 0
\label{eq3} \\
&\sum_{l} \biggl\{ 2 F_l^{\prime}  -
\frac{F_l^{\prime}}{1+F_l^{\prime}/(2l+1)} - \: \bigl( 2 G_l -
\frac{G_l}{1+G_l/(2l+1)} \bigr) \biggr\} \nonumber \\
& \hspace{0.8cm} = \: z^2 \: \frac{16 \, m_{\text{N}} \, k_{\text{F}}
(1+F_1/3)}{4 \pi} \; V_{\text{low k}}
(0,0;\Lambda=k_{\text{F}},{}^1\text{S}_0- {}^3\text{S}_1) \, .
\label{eq4}
\end{align}
We note that the relevant input for the spin-isospin 
Fermi liquid parameters $G_l^\prime$, Eq.
(\ref{eq2}), is the spin averaged s-wave low momentum potential,
whereas the one for the spin Fermi liquid parameters $G_l$,
Eq. (\ref{eq4}), is the difference of the spin singlet and spin
triplet s-wave low momentum potentials. Since the $^3\text{S}_1$
channel is only slightly more attractive than the $^1\text{S}_0$
channel, the right hand side of Eq. (\ref{eq4}) is small.
Consequently, this constraint is not very sensitive to the precise
value of the renormalization factor $z^2$.

The quasiparticle strength $z$ was recently computed in a
self-consistent description of the nucleon spectral functions.
Roth~\cite{zfactor} finds $z=0.76$ at the Fermi surface for
$k_{\text{F}}=1.35 \; \text{fm}^{-1}$.
However, there is a systematic uncertainty on the value of the $z$ factor, 
since the relevance of experimental constraints from (e,e$^\prime$p)
knockout reactions on the jump in the occupation number at the
Fermi surface is questionable. Furnstahl and Hammer~\cite{Furn}
have recently shown that within the rigorous effective field theory
for the interacting dilute Fermi gas the occupation numbers are not 
observable. We use $z=0.8$ and
for the nucleon effective mass at the saturation point we use
$m^{\star}/m=0.72$ corresponding to
\begin{equation}
F_1 = -0.85 .
\end{equation}
The empirical value for the anomalous orbital gyromagnetic ratio
provides a constraint on $F_1^\prime$. For a proton in the Pb
region~\cite{GDR}, $\delta g_l = 0.23 \pm 0.03$. In Fermi liquid
theory~\cite{Migdal} $\delta g_l = (1/3) (F_1^\prime -
F_1)/(1+F_1/3)$, which for $F_1 = -0.85$ yields
\begin{equation}
F_1^{\prime} = 0.14 .
\end{equation}
The incompressibility of nuclear matter is experimentally best
constrained by the isoscalar giant monopole resonance and by
fitting binding energies and the diffuseness of the nuclear
surface. Microscopic calculations by Blaizot {\it et
al.}~\cite{incomp1} and Youngblood {\it et al.}~\cite{incomp2} and
the Thomas-Fermi equation of state of Myers and
Swiatecki~\cite{lidrop} give an incompressibility of $K=230 \pm 20 \;
\text{MeV}$. The Thomas-Fermi equation of state gives a symmetry energy of
$E_{\text{sym}}=32.7 \; \text{MeV}$. The empirical value of the
symmetry energy is limited by various fits to nuclear
masses, resulting in $E_{\text{sym}}=31 \pm 5 \;
\text{MeV}$~\cite{symen}. Thus, we find
\begin{align}
F_0 &= -0.27 \\
F_0^{\prime} &= 0.71 .
\end{align}
In units where $m=1$, the matrix elements
of the low momentum nucleon-nucleon interaction are given
by~\cite{Vlowkflow}
\begin{align}
V_{\text{low k}} (0,0;\Lambda=k_{\text{F}},{}^1\text{S}_0) &= - 1.95 \;
\text{fm} \\
V_{\text{low k}} (0,0;\Lambda=k_{\text{F}},{}^3\text{S}_1) &= - 2.51 \;
\text{fm} .
\end{align}
These are identical for the Bonn-A, Paris, and Argonne-V18
potential as well as a chiral model.

Very similar results are obtained by Feldmeier {\it et
al.}~\cite{UCOM1,UCOM2}, who introduce a unitary correlation operator
including central and tensor correlations. For both the Bonn-A and
Argonne-V18 potentials, they find $V_{\text{UCOM}} \, (0,0;
{}^1\text{S}_0) = - 1.88 \;\text{fm}$ and $V_{\text{UCOM}} \, (0,0;
{}^3\text{S}_1) = - 2.86 \;\text{fm}$. The value in the
$^3\text{S}_1$ channel depends on the range of the tensor
correlations, which for the value quoted here is chosen to
reproduce the d-state admixture when the uncorrelated deuteron
trial wave function contains only an s-wave
component. The dependence on the range of the tensor
correlation corresponds to the cutoff dependence of
$V_{\text{low k}}(0,0;\Lambda, {}^3\text{S}_1)$ discussed in the
introduction.

Epelbaoum {\it et al.}~\cite{Okubo} constructed an effective
potential from a s-wave Malfliet-Tjon type potential. The
transformation method of Okubo used in their work is similar to the
RG decimation employed for $V_{\text{low k}}$. They find $V_{\text{eff}}
\, (0,0; {}^1\text{S}_0) = - 1.94 \; \text{fm}$ for a cutoff of $\Lambda
= 300 \; \text{MeV}$. Their results are in a good agreement
with $V_{\text{low k}}$.

\begin{figure}
\includegraphics[scale=0.375,clip=]{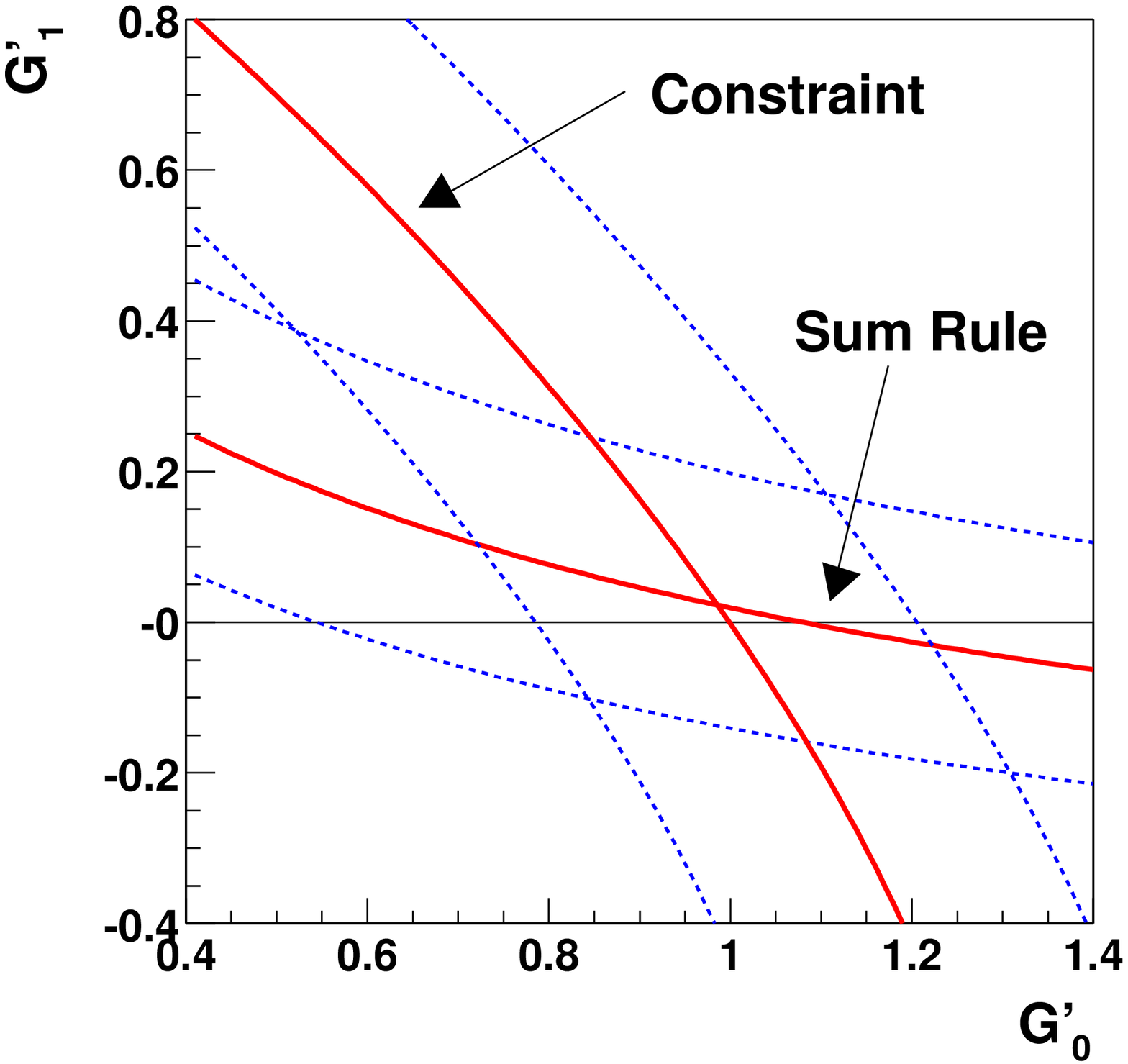}
\hspace{-0.9cm}
\includegraphics[scale=0.375,clip=]{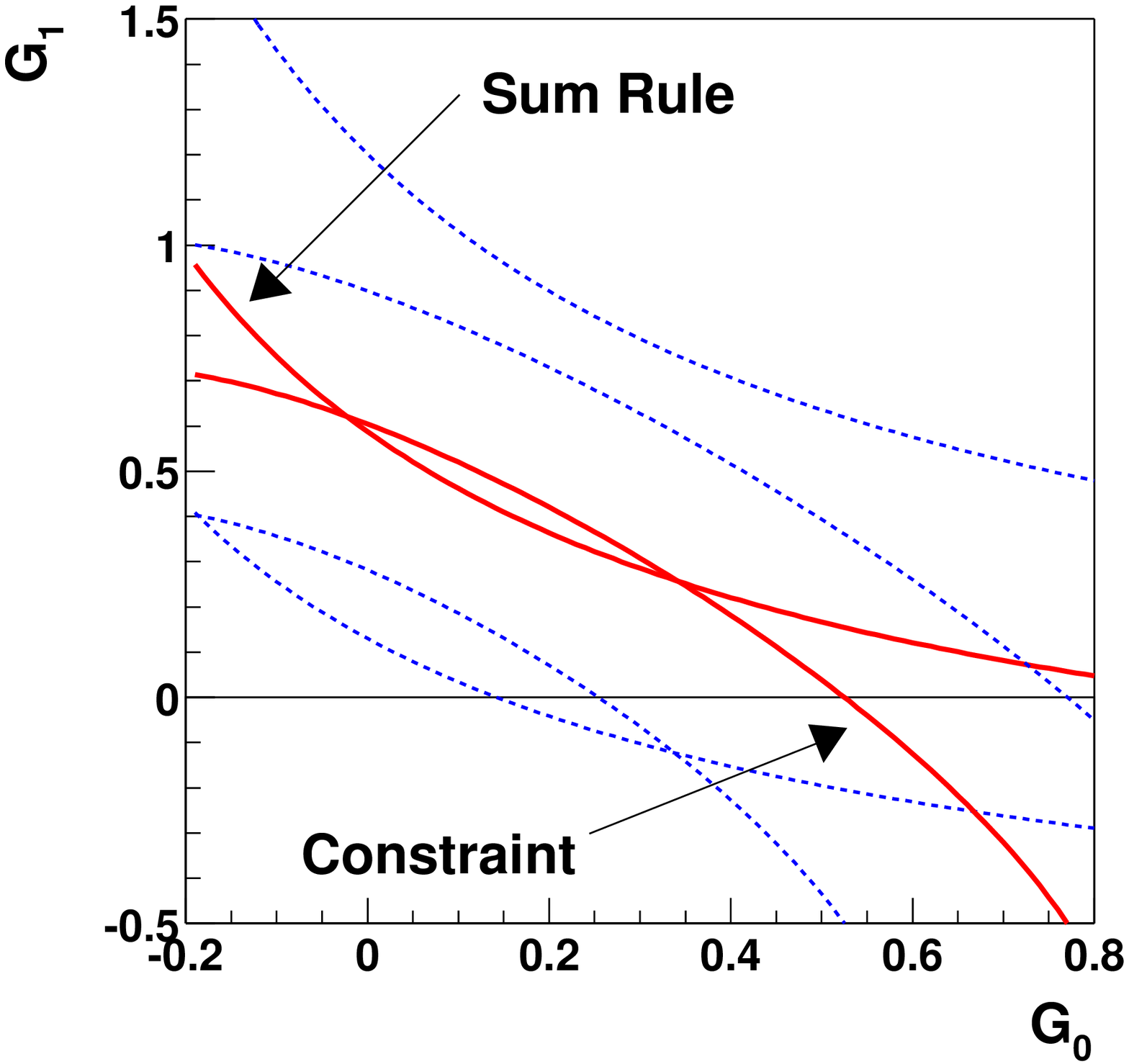}
\caption{The solution for the spin-dependent Fermi liquid parameters
(solid lines) with error bands limited by the dashed lines. Here
the effect of tensor parameters is neglected.}
\label{fitnotensor}
\end{figure}

In Fig.~\ref{fitnotensor} we show the solution to the Eqs.
(\ref{eq1},\ref{eq2},\ref{eq3},\ref{eq4}) without tensor Fermi
liquid parameters. In the error estimates we include the
uncertainties in the input Fermi liquid parameters, the uncertainties
due to the truncation of the Legendre series as well as the
uncertainties in the driving term. The latter include only the
estimated error of the renormalization factor $z$, since
the effects of the neglected higher
order contributions are difficult to appraise. We thus find
\begin{align}
G_0 &= 0.15 \pm 0.3 \\
G_1 &= 0.45 \pm 0.3 \\
G_0^\prime &= 1.0 \pm 0.2 \\
G_1^\prime &= 0 \pm 0.2 .
\end{align}
The relative errors of $G_0$ and $G_1$ are large, because the
corresponding bands are almost parallel. Nevertheless, this
calculation demonstrates that $V_{\text{low k}}$ is a very
promising starting point for calculations of Fermi liquid
parameters.

The mean value of $G_0^\prime=1$ should be confronted with the
experimental constraints imposed by the energy of the
giant Gamow-Teller resonance. Since the
Fermi liquid parameters embody the effective interaction in the
nucleon subspace, the empirical value of $g_{\text{NN}}^\prime$,
obtained in a model that includes $\Delta$-isobar degrees of
freedom, must be corrected for the screening due to $\Delta$-hole
excitations. Including $\Delta$-hole excitations to all orders, we
find
\begin{equation}
G_0^\prime = N(0) \frac{f_{\pi \text{N}
\text{N}}^2}{m_\pi^2}\biggl\{ g_{\text{NN}}^\prime -
\frac{\frac{f^2_{\pi
\text{N} \Delta}}{m_\pi^2} \, g^{\prime 2}_{\text{N} \Delta} \,
\frac{8}{9} \, \frac{\rho_0}{m_{\Delta} - m_{\text{N}}}}{1 + \frac{f^2_{\pi
\text{N} \Delta}}{m_\pi^2} \, g^\prime_{\Delta \Delta} \,
\frac{8}{9} \, \frac{\rho_0}{m_{\Delta} - m_{\text{N}}}} \biggr\} \, ,
\end{equation}
where $\rho_0$ denotes the nuclear matter density.
Furthermore, $g_{\text{N}\text{N}}^\prime$
is the short-range part of the spin-isospin dependent effective
nucleon-nucleon interaction in pionic units, while
$g_{\text{N}\Delta}^\prime$ and $g_{\Delta\Delta}^\prime$ are the
corresponding NN $\rightarrow$ N$\Delta$ and
N$\Delta \rightarrow \Delta$N interaction strengths.
Kawahigashi {\it et al.}~\cite{gp1} find $g_{\text{N}\text{N}}^\prime =
0.6$, $g_{\text{N}\Delta}^\prime=0.3$, while K\"orfgen {\it et
al.}~\cite{gp2,gp3} obtain $g_{\text{N}\Delta}^\prime=0.3$,
$g_{\Delta\Delta}^\prime=0.3$. Using these values, we find $G_0^\prime
= 1.0$ in good agreement with our result. 
The $\Delta$-hole polarization reduces the value of $G_0^\prime$ by
about $10 \%$.

The discussion presented above is easily generalized to include the
effects of the tensor force.
For ${\mathbf p} = {\mathbf p^\prime}$ the tensor components
of the quasiparticle interaction ${\mathcal F}$, Eq. (\ref{LF}), and the
quasiparticle scattering amplitude ${\mathcal A}$ vanish and the
tensor force enters only together with the spin-dependent parameters
in the ${\bm \sigma} \cdot {\bm \sigma^{\prime}}$ and ${\bm \tau} \cdot {\bm
\tau^{\prime}} \: {\bm \sigma} \cdot {\bm \sigma^{\prime}}$ components
of the scattering amplitude~\cite{BD}. The coupling of spin and
Landau $l$ to good total angular momentum $J$ was carried out
by B\"ackman {\it et al.}~\cite{tensor}. 
We use the tensor Fermi liquid parameters
obtained from a $G$ matrix calculation using Reid's soft core
potential for $l
\leq 4$ and from the one pion exchange potential for higher $l$
(see Table 2 of~\cite{tensor}). To account for the effects of
tensor forces,  we replace the spin-dependent parameters of the
scattering amplitude $C_l=G_l/(1+G_l/(2l+1))$ and $C_l^\prime =
G_l^\prime/(1+G_l^\prime/(2l+1))$ in the constraints up to $l=4$
with the corresponding expressions including tensor
interactions~\cite{BD}. Since the expansion of
the tensor interaction in Landau $l$ is poorly convergent, we
include terms up to $l=4$. We have checked that the contributions
of higher $l$ are negligible. We note that the tensor parameters 
of Ref.~\cite{tensor} are given for $z=1$ and $m^\star/m=1$.
Consequently, these parameters should be reduced by the factor
$z^2 \,\frac{m^\star}{m}$.

\begin{figure}
\includegraphics[scale=0.375,clip=]{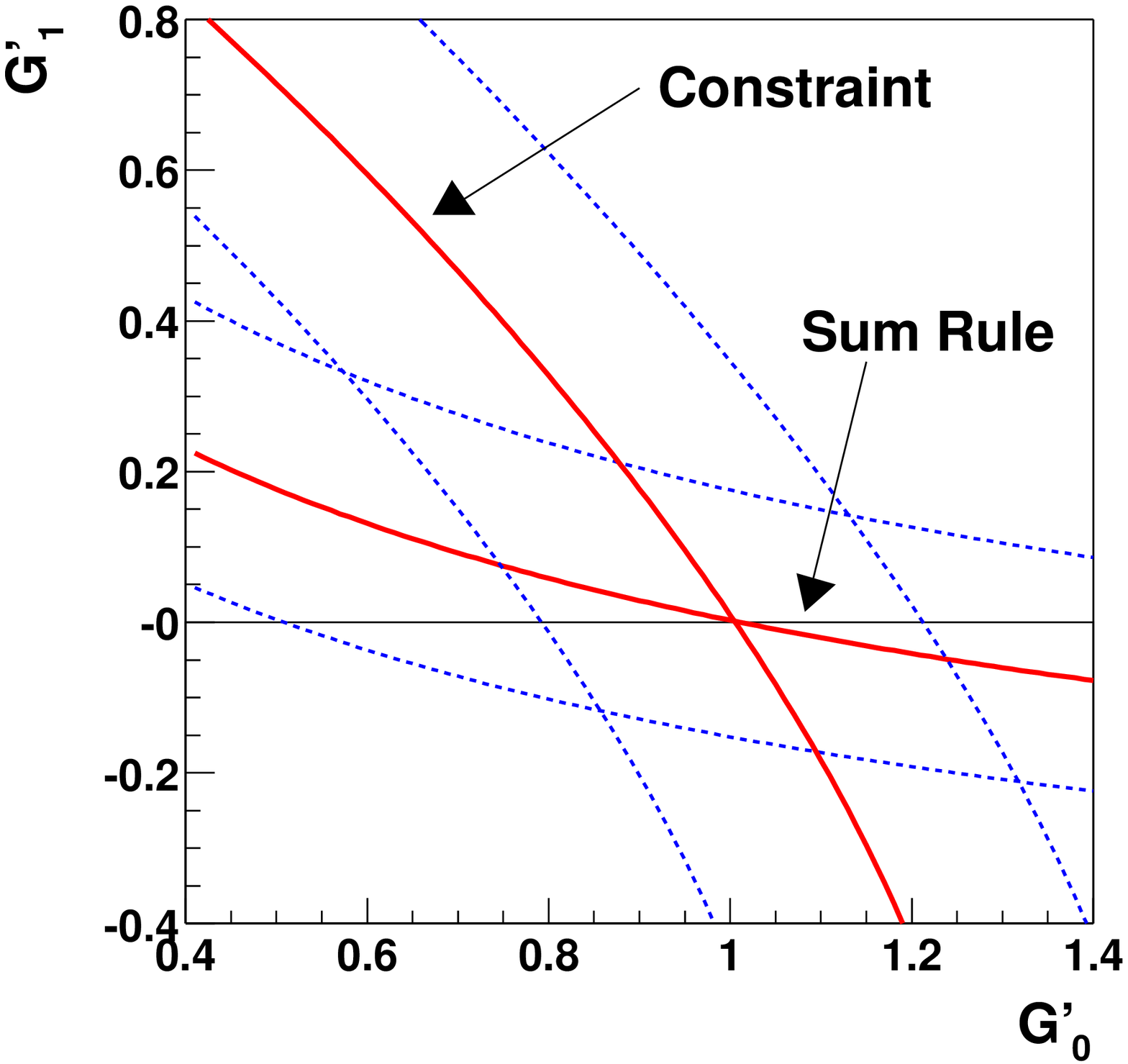}
\hspace{-0.9cm}
\includegraphics[scale=0.375,clip=]{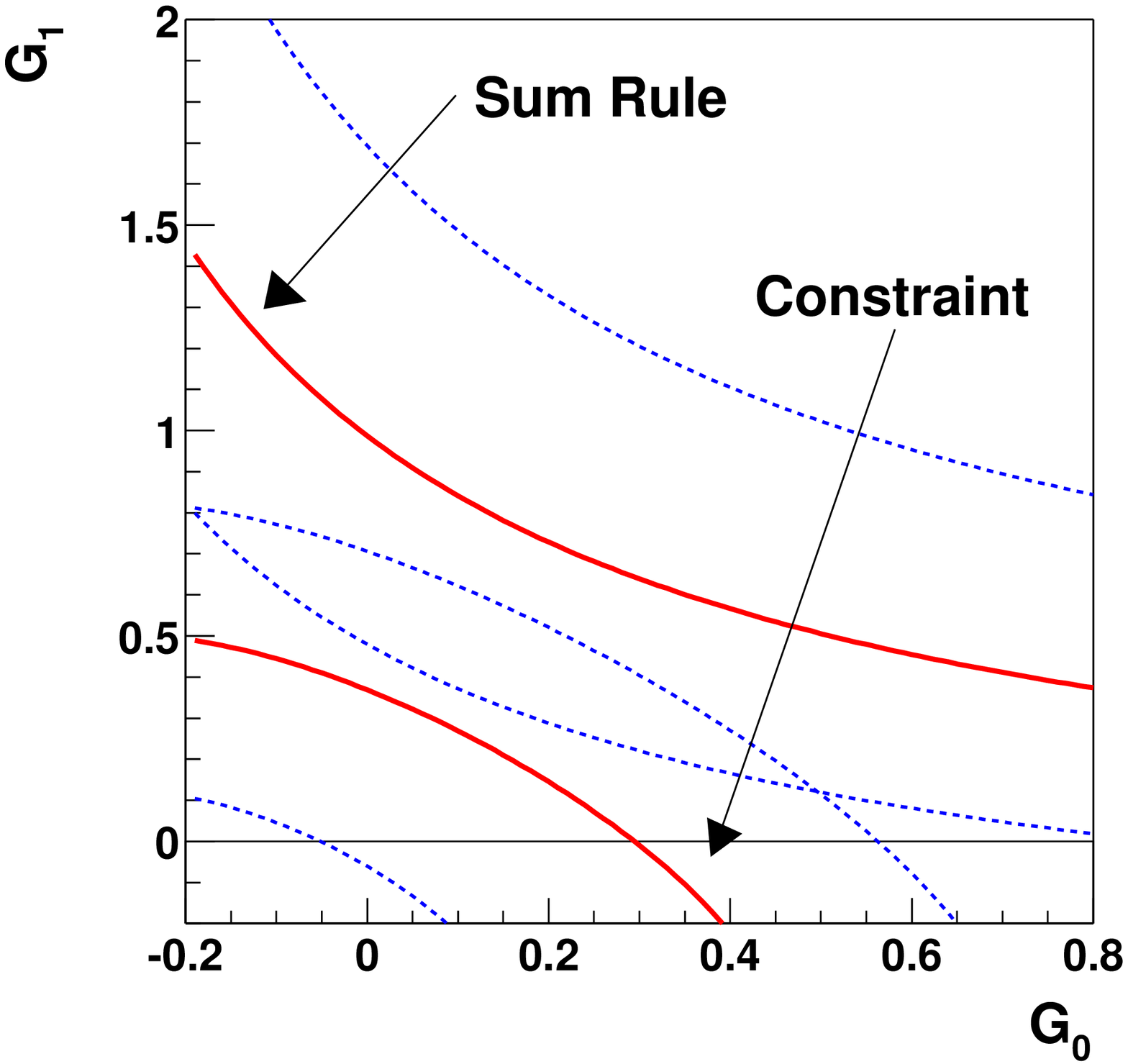}
\caption{The same as Fig.~1 but including tensor interactions.}
\label{fitwithtensor}
\end{figure}

In Fig.~\ref{fitwithtensor} we show the solution to the Eqs.
(\ref{eq1},\ref{eq2},\ref{eq3},\ref{eq4}) including tensor Fermi
liquid parameters. We have not included errors
for the tensor parameters. Since the isospin tensor parameters
$H_l^\prime$ are small (one pion and one rho exchange yields
$H_l^\prime = - H_l/3$~\cite{bigreport}), the solution for
$G_0^\prime$ and $G_1^\prime$ in the spin-isospin sector is
basically unaffected by the presence of tensor interactions.
However, the solution for the spin Fermi liquid parameters $G_0$
and $G_1$ is strongly modified. In fact,
the error bands overlap only in a small region, when we include
tensor interactions. The reason is that in this sector the tensor
parameters of Ref.~\cite{tensor} are quite large. We note that this
may change, when the contribution of the induced interaction to the
tensor parameters is included. In order to illustrate the possible
effects of this type, we compute the leading contribution to the
tensor Fermi liquid parameters from the one bubble polarization in
the induced interaction using the one pion exchange interaction.
The lowest order contribution to the tensor Fermi liquid parameters
$H_l$ from the one pion exchange driving term, Eq. (\ref{OPE}), is
given by
\begin{equation}
H(\theta) = N(0) \: z^2 \frac{f^2}{3 \: m_\pi^2} \: \frac{3}{2} \:
\frac{k_{\text{F}}^2}{({\mathbf p} -{\mathbf p^\prime})^2 +m^2_\pi}
\, .
\end{equation}
The dominant tensor contribution from the one bubble term in the
induced interaction is obtained by
employing the direct tensor part $\pi_{\text{T}}$ of the one pion
exchange potential as vertices in the induced interaction. This
corresponds to the first and in part the third and the last
diagrams of Eq. (\ref{diag2ndorder}). More explicitly, we compute the
diagrams
\begin{equation}
\parbox[c][2.8cm]{2.7cm}{\begin{fmfgraph*}(25,25)
\fmfleft{p1,pp1}
\fmfright{p2,pp2}
\fmfforce{(0,0)}{p1}
\fmfforce{(w,0)}{p2}
\fmfforce{(0,h)}{pp1}
\fmfforce{(w,h)}{pp2}
\fmfforce{(0.1w,0.3h)}{v1}
\fmfforce{(0.9w,0.7h)}{v2}
\fmfforce{(0.5w,0.3h)}{v3}
\fmfforce{(0.5w,0.7h)}{v4}
\fmf{fermion}{p1,v1,pp1}
\fmf{dashes,label=\large$\pi_{\text{T}}$}{v1,v3}
\fmf{dashes,label=\large$\pi_{\text{T}}$}{v2,v4}
\fmf{fermion}{pp2,v2,p2}
\fmf{fermion,left=0.5,tension=0.3}{v3,v4}
\fmf{fermion,left=0.5,tension=0.3}{v4,v3}
\end{fmfgraph*}} \hspace{0.1cm} + \hspace{0.2cm}
\parbox[c][2.8cm]{2.7cm}{\begin{fmfgraph*}(25,25)
\fmfleft{p1,pp1}
\fmfright{p2,pp2}
\fmfforce{(0,0)}{p1}
\fmfforce{(w,0)}{p2}
\fmfforce{(0,h)}{pp1}
\fmfforce{(w,h)}{pp2}
\fmfforce{(0.05w,0.25h)}{v1}
\fmfforce{(0.95w,0.75h)}{v2}
\fmfforce{(0.5w,0.25h)}{v3}
\fmfforce{(0.5w,0.75h)}{v4}
\fmf{fermion}{p1,v1,pp1}
\fmf{dashes,label=\large$\pi_{\text{T}}$}{v1,v3}
\fmf{fermion}{pp2,v2,p2}
\fmf{fermion,left=0.4}{v3,v4}
\fmf{fermion,left=0.4}{v4,v3}
\fmfpoly{shaded,smooth}{v2,v5,v4,v6}
\fmfforce{(0.725w,0.85h)}{v5}
\fmfforce{(0.725w,0.65h)}{v6}
\fmfv{label=\large$G^\prime$,label.angle=80}{v5}
\end{fmfgraph*}} \hspace{0.1cm} + \hspace{0.2cm}
\parbox[c][2.8cm]{2.7cm}{\begin{fmfgraph*}(25,25)
\fmfleft{p1,pp1}
\fmfright{p2,pp2}
\fmfforce{(0,0)}{p1}
\fmfforce{(w,0)}{p2}
\fmfforce{(0,h)}{pp1}
\fmfforce{(w,h)}{pp2}
\fmfforce{(0.05w,0.25h)}{v1}
\fmfforce{(0.95w,0.75h)}{v2}
\fmfforce{(0.5w,0.25h)}{v3}
\fmfforce{(0.5w,0.75h)}{v4}
\fmf{fermion}{p1,v1,pp1}
\fmf{dashes,label=\large$\pi_{\text{T}}$}{v2,v4}
\fmf{fermion}{pp2,v2,p2}
\fmf{fermion,left=0.4}{v3,v4}
\fmf{fermion,left=0.4}{v4,v3}
\fmfpoly{shaded,smooth}{v1,v5,v3,v6}
\fmfforce{(0.275w,0.15h)}{v5}
\fmfforce{(0.275w,0.35h)}{v6}
\fmfv{label=\large$G^\prime$,label.angle=-80}{v5}
\end{fmfgraph*}} .
\end{equation}
For the long range part of $G^\prime$ we include the momentum
dependence by splitting the interaction into a one pion exchange piece
and a short ranged piece~\footnote{This is justified since the
$q$ dependence of the exchanged heavy mesons is weak.}:
\begin{equation}
G^\prime = N(0) \: z^2 \frac{f^2}{3 \: m_\pi^2} \: \bigl(
\frac{m_\pi^2}{{\mathbf q}^2 + m^2_\pi} + \Delta g^\prime \bigr) \, .
\end{equation}
Using Eqs. (\ref{OPE}) and (\ref{II2}) we then find
\begin{align}
& \Gamma_{{\bm \sigma} \cdot {\bm \sigma^{\prime}}, {\bm \tau} \cdot
{\bm \tau^{\prime}}}^{\text{ind (2) dir. OPE}} (p,p^\prime)
= - \frac{1}{4} \: (1+{\bm \sigma} \cdot {\bm \sigma^{\prime}}) \: (3-{\bm
\tau} \cdot {\bm \tau^{\prime}}) \: N(0) \: z^2 \: \bigl( \frac{f^2}{3
\: m_\pi^2} \bigr)^2 \nonumber \\[0.5mm]
& \: \times \frac{1}{2} \:
\text{Tr}_{{\bm \sigma^{\prime\prime}}} \: \int \frac{d \Omega_{{\mathbf
p^{\prime\prime}}}}{4 \pi}
\: \: ({\mathbf
p}-{\mathbf p^{\prime}})^2 \frac{S_{1 2^{\prime\prime}}
({{\mathbf p} - {\mathbf
p^{\prime}}})}{({\mathbf p} -{\mathbf p^{\prime}})^2 +m^2_\pi} \:
\biggl\{ ({\mathbf
p}-{\mathbf p^\prime})^2 \frac{S_{2^{\prime\prime} 2} ({{\mathbf p} - {\mathbf
p^\prime}})}{({\mathbf p} -{\mathbf p^\prime})^2 +m^2_\pi} \nonumber \\[1mm]
& \quad - 2 \,
\frac{m^2_\pi \: {\bm \sigma^{\prime\prime}} \cdot {\bm
\sigma^{\prime}}}{({\mathbf p} -{\mathbf
p^\prime})^2+m^2_\pi} - 6 \, \Delta g^\prime \biggr\} \:
\frac{v_{\text{F}} \: \hat{{\mathbf p}}^{\prime\prime} \cdot ({\mathbf
p} - {\mathbf p^\prime})}{\varepsilon - \varepsilon^\prime -
v_{\text{F}} \: \hat{{\mathbf p}}^{\prime\prime} \cdot ({\mathbf p} -
{\mathbf p^\prime})} \, .
\end{align}
The integral over $\Omega_{{\mathbf p^{\prime\prime}}}$ yields the
Lindhard function $\displaystyle -\alpha_0
(q^\prime/k_{\text{F}})$. By exploiting the following identities
for the tensor operator,
\begin{align}
\frac{1}{2} \:
\text{Tr}_{{\bm \sigma^{\prime\prime}}} \: S_{1 2^{\prime\prime}}
({{\mathbf p} - {\mathbf p^{\prime}}}) \: S_{2^{\prime\prime} 2}
({{\mathbf p} - {\mathbf p^\prime}}) & = S_{1 2}
({{\mathbf p} - {\mathbf p^{\prime}}}) + 2 \: {\bm \sigma}
\cdot {\bm \sigma^{\prime}} \\
\frac{1}{2} \: \text{Tr}_{{\bm \sigma^{\prime\prime}}} \: S_{1
2^{\prime\prime}} ({{\mathbf p} - {\mathbf p^{\prime}}}) \: {\bm
\sigma^{\prime\prime}} \cdot {\bm \sigma^{\prime}} & = S_{1 2}
({{\mathbf p} - {\mathbf p^{\prime}}}) \\ \frac{1}{2} \:
(1+{\bm \sigma} \cdot {\bm \sigma^{\prime}}) \: S_{1 2}
({{\mathbf p} - {\mathbf p^{\prime}}}) & = S_{1 2}
({{\mathbf p} - {\mathbf p^{\prime}}}) \, ,
\end{align}
we finally arrive at the second order correction to the tensor
Fermi liquid parameters
\begin{multline}
\Delta H (\theta) = H(\theta) \: N(0) \: z^2 \frac{f^2}{3 \: m_\pi^2} \:
\alpha_0 (q^\prime/k_{\text{F}}) \\ \times \biggl\{
\frac{({\mathbf p} -{\mathbf
p^\prime})^2}{({\mathbf p} -{\mathbf p^\prime})^2 +m^2_\pi} - 2 \,
\frac{m^2_\pi}{({\mathbf p} -{\mathbf p^\prime})^2 +m^2_\pi} - 6 \,
\Delta g^\prime \biggr\} \, .
\label{DeltaH}
\end{multline}
In order to reproduce the empirical value for $G_0^\prime=1$ with
the direct one pion exchange contribution plus $\Delta g^\prime$,
we need $\Delta g^\prime=0.5$~\footnote{Note that the value of
$g_{\text{NN}}^\prime$ here is larger than in~\cite{gp1}, because we
use $z < 1$. The physics is determined by $G_0^\prime$, not by
$g_{\text{NN}}^\prime$.}. The resulting corrections to the tensor
Fermi liquid parameters $H_0 = 0.35$ and $H_1
= 0.43$ are $\Delta H_0 = -0.40$ and
$\Delta H_1 = -0.69$. Thus, we find that the induced interaction
tends to reduce the tensor Fermi liquid parameters, in agreement
with the results of Dickhoff {\it et al.}~\cite{Dickhoff}. The very
large effects show that the tensor interactions must be treated
self consistently within the induced interaction. Finally, we note
that about $60 \%$ of the left hand side of Eq. (\ref{eq2}) is due
to the Landau parameter $G_0^\prime$. Thus, there is a close
connection between the spin-averaged s-wave low momentum
interaction $V_{\text{low k}} (0,0;\Lambda=k_{\text{F}})$ and the
local spin-isospin dependent part of the quasiparticle
interaction.

\section{Summary and Conclusions}

In this paper we presented two new algebraic constraints that
relate the Landau Fermi liquid parameters in nuclear matter to the
driving term of the induced interaction. By identifying the driving
term with the  s-wave low momentum nucleon-nucleon interaction
$V_{\text{low k}}$ at $\Lambda = k_{\text{F}}$, including some
straightforward in-medium effects, we obtained an intriguing
relation between the effective interaction {\it in vacuum} and {\it in nuclear
matter}.

The resulting constraints on the Fermi liquid parameters were used
in conjunction with the Pauli principle sum rules to compute the
major spin dependent parameters, given the phenomenological values
for the spin independent parameters. We find good
agreement with empirically determined parameters. The present
calculation indicates that a good approximation to the driving term 
of the induced interaction can be obtained from $V_{\text{low k}}$ in a
straightforward manner, by including minimal in-medium corrections,
the wave-function renormalization factors and the nucleon effective
mass in the density of states. A full calculation of the induced
interaction, including a self-consistent treatment would be needed
to firmly establish this identification. In such a calculation, the
spin, isospin and velocity dependence of $V_{\text{low k}}$ would be reflected 
in the corresponding Fermi liquid parameters. A
comparison with empirical parameters would then provide a test of e.g.
the velocity dependence of the low momentum nucleon-nucleon interaction
$V_{\text{low k}}$. In $V_{\text{low k}}$, the role of the (local) short range
repulsion of the bare interactions is taken over by a non-locality,
which interpolates between a weak repulsion at low energies and a
stronger one at higher energies.

The effects of tensor forces are also studied, using the tensor
parameters obtained in a $G$ matrix calculation. We find a fairly
large effect of the tensor force on the isoscalar spin dependent
parameters. However, as indicated by a simple estimate, this effect
will probably be reduced when the tensor parameters are computed
self consistently by including the tensor force in the induced
interaction.

Moreover, we derive the flow equations for the renormalization group decimation
of the quasiparticle scattering amplitude and the quasiparticle
interaction in the two particle-hole channels starting from the
induced interaction. A solution of these equations would provide
the scattering amplitude also for non-forward scattering, which is
of high interest for the calculation of superfluid gaps and
transport processes, e.g. in neutron star interiors. In condensed
matter systems an ab initio RG analysis of this type~\cite{Metzner} applied to
the 2D Hubbard model has successfully established the existence of d-wave
superconductivity. The RG equations for the quasiparticle scattering 
amplitude and the quasiparticle interaction
we obtained from the induced interaction are nonperturbative. 
Existing RG studies in Fermi systems have been restricted to one-loop 
approximations. 

\begin{ack}
We thank Scott Bogner and Tom Kuo for helpful discussions. AS acknowledges
the kind hospitality of the Theory Group at GSI, where
part of this work was carried out. The work of AS and GEB was
supported by the US-DOE grant No. DE-FG02-88ER40388.
\end{ack}

\end{fmffile}
\end{document}